\documentclass[lettersize,journal]{IEEEtran}
\usepackage{amsmath,amsfonts}
\usepackage{algorithmic}
\usepackage{algorithm}
\usepackage{array}
\usepackage[caption=false,labelformat=simple]{subfig}

\usepackage{textcomp}
\usepackage{stfloats}
\usepackage{url}
\usepackage{verbatim}
\usepackage{graphicx}
\usepackage{cite}
\usepackage{booktabs}
\usepackage[hidelinks]{hyperref} 
\usepackage{makecell}
\usepackage{xcolor}
\usepackage[switch]{lineno}
\newcommand\TS{\rule{0pt}{2.6ex}}

\usepackage{array, multirow}
\begin{document}
\title{An Overview of Algorithms for Contactless Cardiac Feature Extraction from Radar Signals: Advances and Challenges}

\author{Yuanyuan~Zhang, Rui~Yang, \IEEEmembership{Member,~IEEE}, Yutao~Yue, \\ Eng~Gee~Lim, \IEEEmembership{Senior~Member,~IEEE}, Zidong~Wang, \IEEEmembership{Fellow,~IEEE}
\thanks{This work is partially supported by the National Natural Science Foundation of China (61603223), the Jiangsu Provincial Qinglan Project (2021), the Research Development Fund of XJTLU (RDF-20-01-18), and the Suzhou Science and Technology Programme (SYG202106). This work is also financially supported by Jiangsu Industrial Technology Research Institute (JITRI) and Wuxi National Hi-Tech District (WND). \textit{(Corresponding authors: Rui Yang, Yutao Yue.)}}
\thanks{Yuanyuan Zhang is with the School of Advanced Technology, Xi’an Jiaotong-Liverpool University, Suzhou, 215123, China, the Department of Electrical Engineering and Electronics, University of Liverpool, Liverpool, L69 3GJ, United Kingdom, and also with the Institute of Deep Perception Technology, JITRI, Wuxi, 214000, China (email: Yuanyuan.Zhang16@student.xjtlu.edu.cn).}
\thanks{Rui Yang and Eng Gee Lim are with the School of Advanced Technology, Xi’an Jiaotong-Liverpool University, Suzhou, 215123, China.}
\thanks{Yutao Yue is with the Institute of Deep Perception Technology, JITRI, Wuxi, 214000, China,
the XJTLU-JITRI Academy of Industrial Technology, Xi'an Jiaotong-Liverpool University, Suzhou, 215123, China, and also with the Department of Mathematical Sciences, University of Liverpool, Liverpool, L69 7ZX, UK.}
\thanks{Zidong Wang is with the Department of Computer Science, Brunel University London, Uxbridge, UB8 3PH, United Kingdom.}}

\maketitle

\begin{abstract}
Contactless cardiac monitoring has vast potential to replace contact-based monitoring in various future scenarios such as smart home and in-cabin monitoring. Various contactless sensors can be potentially implemented for cardiac monitoring, such as cameras, acoustic sensors, Wi-Fi routers and radars. Among all these sensors, radar could achieve unobtrusive monitoring with high accuracy and robustness at the same time. The research about radar-based cardiac monitoring can be generally divided into the radar architecture design and signal-processing parts, where the former has been thoroughly reviewed in the literature but not the latter. To the best of the author’s knowledge, this is the first review paper that focuses on elaborating the algorithms for extracting cardiac features from the received radar signal. In addition, a new taxonomy is proposed to reveal the core feature of each algorithm, with the pros and cons evaluated in detail. Furthermore, the public datasets containing the received radar signal and ground-truth cardiac feature signal are listed with detailed configurations, and the corresponding evaluations may help the researchers select the suitable dataset. At last, several unsolved challenges and future directions are suggested and discussed in detail to encourage future research on solving the main obstacles in this field. In summary, this review can be served as a guide for researchers and practitioners to quickly understand the research trend and recent development of the cardiac feature extraction algorithms, and it is worth further investigating the relative area based on the proposed challenges and future directions.
\end{abstract}


\section{Introduction}
Cardiac monitoring is a faithful diagnostic tool widely used in clinical medicine to measure various cardiac features (e.g., heart rate (HR), beat-to-beat interval (BBI), heart rate variability (HRV), electrocardiogram (ECG), seismocardiography (SCG))~\cite{krittanawong2021integration}. Some of the cardiac features, such as ECG and SCG, describe the electrical and mechanical activities of the heart respectively, while other features, such as BBI and HRV, can reveal some cardiac diseases in terms of the variations of values~\cite{chen2018arrhythmia}. Various commercial products have already achieved a high-accuracy measurement of cardiac features with the aid of electrode patches or wearable devices~\cite{chen2021contactless}. Despite the efficiency, these contact-based measurements may be unfriendly to patients with burn or dermatosis~\cite{ye2021spectral} and unsuitable for long-term monitoring~\cite{singh2020multi}, bringing uncomfortable feelings to the patients~\cite{jia2018wifind}. Therefore, contactless cardiac monitoring techniques are imperative to be developed and can be deployed in the smart home~\cite{adib2015smart}, driver monitoring system~\cite{kang2013various} and post-disaster search in the future~\cite{shen2018respiration}.

Various contactless sensors can be utilized for cardiac monitoring according to the information received. For example, optical/thermal cameras sense the skin colour/temperature variation caused by heartbeats~\cite{chen2018video,chian2022vital}, acoustic sensors can monitor the heart sound~\cite{xu2020leveraging}, and Wi-Fi routers extract cardiac features from the channel state information~\cite{jia2018wifind}. However, the aforementioned sensors may be blamed for privacy issue~\cite{feng2021multitarget}, low accuracy~\cite{nirmal2021deep} or vulnerability to the changing environment (e.g., light conditions or temperature variations)~\cite{xia2021radar}. In contrast, radar senses the ambient environment through reflected signals mixed by heart vibration, chest wall displacement (induced by respiratory and cardiac activities) and all kinds of ambient noises~\cite{wang2020remote}, requiring proper algorithms to further extract the latent cardiac features. Additionally, compared with cameras, Wi-Fi routers and acoustic sensors, radar signal propagation is neither vulnerable to the illumination/temperature/sound variations nor privacy-intrusive. With advanced signal-processing algorithms, radar-based cardiac monitoring is promising to realize unobtrusive cardiac monitoring in most scenarios.

\begin{figure*}[tb] 
    \centering 
    \includegraphics[width=1.9\columnwidth]{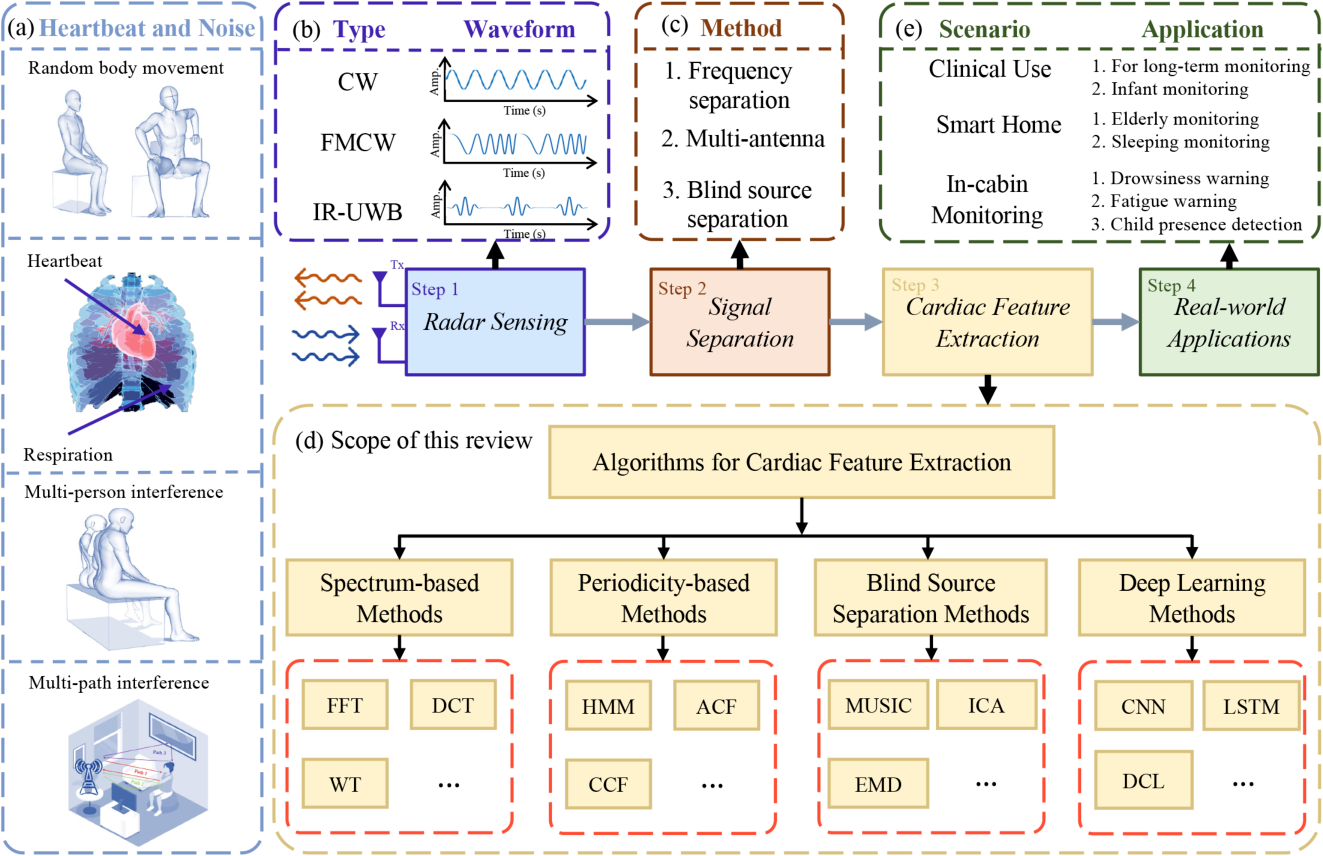}
    \caption{Overview of radar-based cardiac monitoring: (a) target heartbeat signal and other noises; (b) main radar types: continuous wave (CW) radar, frequency modulated continuous wave (FMCW) radar and impulse-radio ultra wide band (IR-UWB) radar; (c) examples of algorithms for multi-person separation~\cite{islam2022contactless}; (d) algorithms for cardiac feature extraction; (e) the applicable scenarios for contactless cardiac monitoring.}
    \label{fig:pipeline} 
\end{figure*}

The overview of the radar-based cardiac monitoring process and the downstream applications are shown in Figure~\ref{fig:pipeline}. To measure the target cardiac features, the first step is to transmit the radio frequency signals with certain waveforms to sense the ambient environment as illustrated in Figure~\ref{fig:pipeline}(b). After reflection, the received signals contain not only the target signal (vibration caused by heartbeats) but also other noise signals such as random body movement (RBM), respiration and multi-person or multi-path interferences, as shown in Figure~\ref{fig:pipeline}(a). The second step aims to isolate the signal reflected by a single human body from the background clutter or the interference from neighbours using specific methods, as shown in Figure~\ref{fig:pipeline}(c)~\cite{islam2022contactless}. The third step is to extract the cardiac features from the reflected signals of a single person using various algorithms, with detailed classification provided in Figure~\ref{fig:pipeline}(d). The last step is to analyse the obtained cardiac features for specific downstream applications as shown in Figure~\ref{fig:pipeline}(e). 

Based on the literature review, early studies mainly focused on the first step to improve radar front-end designs (e.g., self-injection-locked radar~\cite{wang2010novel}), modulation techniques (e.g., in-phase/quadrature modulation~\cite{droitcour2004range}), phase unwrapping methods (e.g., arctangent demodulation~\cite{wang2020remote}) and so on. However, the cardiac feature extraction step was normally performed by conventional algorithms, such as fast Fourier transform (FFT) and band-pass filter (BPF), with limited performance in accuracy and noise mitigation. In recent years, reliable commercial radar platforms have simplified the radar configuration setup and enabled researchers to develop advanced algorithms for cardiac feature extraction from received radar signals with research interests and relevant work concluded in Table~\ref{tab:interest}.   

\begin{table*}[tb]
    \centering
    \caption{Research interests with explanations and relevant work}
      \begin{tabular}{ p{0.56\columnwidth} p{1.4\columnwidth} }
      \toprule
      \textbf{Research Interests with Relevant Work} & \textbf{Explanations} \\
      \toprule
      \textbf{To extract fine-grained cardiac features} \quad \cite{chen2021contactless}, \cite{ha2020contactless}, \cite{ji2022rbhhm}, \cite{ha2021wistress}, \cite{yamamoto2020ecg} &  Besides the coarse measurements such as HR and BBI, the fine-grained cardiac features such as ECG or SCG measurements can be reconstructed from radar signals to describe the electrical and mechanical activities of the heart, enabling the evaluation of implicit cardiac event segmentation~\cite{ha2020contactless}, situational awareness~\cite{chen2021contactless} and mental status~\cite{ha2021wistress}. However, extracting fine-grained cardiac features requires a complex mathematical model for the subtle cardiac mechanical or electrical activities (e.g., valve opening~\cite{ha2020contactless} or ventricular depolarization~\cite{ji2022rbhhm}), instead of only detecting the most intensive heartbeat for coarse measurements.\\
      \hline
      \TS
      \textbf{To mitigate real-world noise} \quad \quad \quad \quad \quad \cite{wang2020remote}, \cite{mercuri2021enabling}, \cite{zhang2021mutual}, \cite{chen2021movi}, \cite{lee2016tracking}, \cite{gong2021rf}, \cite{tomii2015heartbeat}, \cite{zhang2020health}, \cite{mogi2017heartbeat}, \cite{zhao2016emotion}, \cite{lv2018doppler}, \cite{diraco2017radar}, \cite{shyu2018detection}, \cite{ye2018stochastic}, \cite{ye2018robust}, \cite{ye2019blind}, \cite{xiong2017accurate}, \cite{xiong2020differential}, \cite{tu2015fast} & The cardiac features can be drowned out by the real-world noises such as background clutter~\cite{kwon2021attention}, high-order respiration harmonics~\cite{wang2020remote}, RBM~\cite{chen2021movi}, multi-path interference~\cite{mercuri2021enabling} and car vibrations~\cite{lee2016tracking}. Many researchers focus on designing algorithms to reduce these noises and achieve noise-robust monitoring in real-world applications. However, only the distortion caused by respiration and slight RBM can be well-addressed currently~\cite{gong2021rf}, while the main unsolved challenges include the mitigation of consistent body movement~\cite{chen2021movi}, modelling of mutual-radar or multi-path interference~\cite{mercuri2021enabling,zhang2021mutual} and so on. \\
      \hline
      \TS
      \textbf{To realize multi-person monitoring} \quad \quad \quad \cite{feng2021multitarget}, \cite{islam2022contactless}, \cite{nosrati2019concurrent}, \cite{mercuri2019vital},\cite{wang2020multiple} &  To enable complex applications with the requirement of multi-person cardiac monitoring such as smart home or clinical monitoring~\cite{mercuri2019vital}, most researchers either leverage the 3D-modelling ability of certain types of radars (e.g., frequency modulated continuous wave (FMCW) radar) to isolate the reflections coming from different range bins~\cite{adib20143d}, or apply advanced antenna design, such as phased-array antenna or digital beamforming, to separate different people from different angles~\cite{feng2021multitarget}. However, the resolution and the robustness of the current multi-person monitoring techniques need to be improved to realize the monitoring of closely adjacent or moving people. \\ 
      \hline
      \TS
      \textbf{To enable multi-radar/sensor monitoring} \quad \cite{ren2021vital}, \cite{gravina2017multi}, \cite{8281483}, \cite{bruser2015ambient},\cite{wang2022multisensor} & Multi-radar and multi-sensor fusion~\cite{bai2018fusion} leverage information provided by extra radars/sensors to locate the body~\cite{gupta2022automatic}, eliminate RBM~\cite{gu2013hybrid}, realize any-orientation monitoring~\cite{ren2021vital} and so on. However, the current multi-radar monitoring system still requires deploying the radars in particular locations~\cite{ren2021vital}, and the multi-sensor data fusion requires specific signal-processing algorithms different from single-radar-based cardiac monitoring to fuse the features extracted from different data modalities~\cite{gravina2017multi}.\\
      \bottomrule
      \end{tabular}%
    \label{tab:interest}%
\end{table*}

Recent research on the interests in Table~\ref{tab:interest} has yielded extensive algorithms for cardiac feature extraction from radar signals, whereas none of the existing reviews systematically concluded the recent advancements in this area because most effective algorithms (especially the deep learning algorithms) were proposed only in recent three years. For the existing reviews, Li \textit{et al.}~\cite{li2013review} mainly summarized the development of radio frequency front-end architectures and baseband signal processing methods before 2013. Obadi \textit{et al.}~\cite{obadi2021survey} first compared the designs of different radar types and then concluded the algorithms implemented on FPGA board, whereas the limited computational resource restricts the complexity of the reviewed algorithms. Islam \textit{et al.}~\cite{islam2022contactless} mainly introduced the methods for multi-person separation and also mentioned several conventional methods for cardiac feature extraction. Singh \textit{et al.}~\cite{singh2020multi} listed the main challenges in radar-based cardiac monitoring, but mainly focused on the radar design and phase unwrapping algorithms. In~\cite{nirmal2021deep}, several deep learning-based algorithms applied for vital sign extraction in recent years were introduced, but many conventional algorithms are still required to be discussed for a systematic review. In summary, the existing reviews have covered multiple topics in radar-based cardiac monitoring, such as radar architecture design, baseband radar signal processing (e.g., phase unwrapping algorithms) and multi-person separation techniques. To expand the scope of the reviewed topics, this work aims to thoroughly review the algorithms for cardiac feature extraction to help new researchers obtain a general impression of related algorithms, current challenges and future directions. The contributions of this review are listed as follows:
\begin{itemize}
    \item To the best of the authors' knowledge, this is the first work that elaborates the algorithms for radar-based cardiac feature extraction, especially the deep learning algorithms.
    \item Based on the core principle of each algorithm, a new taxonomy is proposed as shown in Figure~\ref{fig:pipeline}(d), to help researchers and practitioners make a quick impression on the methods applied in radar-based cardiac feature extraction. In addition, the general principles for the algorithms are also detailed with their pros, cons and future improvements.
    \item This review concludes and evaluates the public datasets that contain synchronized cardiac features and radar signals, with advice on the dataset usage and design in the future.
\end{itemize}

The rest of the review is organized as follows. Section~\ref{sec:prim} introduces the background of radar-based cardiac monitoring with the trend of radar usage. Section~\ref{sec:algos} first explains the proposed taxonomy and then elaborates all the algorithms with their applications mentioned in the literature. The public datasets are listed and evaluated in Section~\ref{sec:dataset}. In Section~\ref{sec:candf}, we provide insight into the challenges and future research directions. Section~\ref{sec:conclusions} concludes this review paper.

\section{Background Knowledge of Radar}\label{sec:prim}
Radar refers to the radio detection and ranging system that transmits the electromagnetic wave to sense the ambient environment and is originally used for military detection of large objects such as aircraft. In 1975, Lin~\cite{lin1975noninvasive} performed the world's first respiration monitoring using radar by measuring chest wall displacement, and such displacement can be reckoned to be a large-scale vibration with higher amplitude than heart vibration~\cite{shen2018respiration}. In recent decades, with advanced radar architecture and signal-processing algorithms, researchers have been able to extend radar-based respiration monitoring to cardiac monitoring, because the underlying principle is always to reconstruct the vital vibrations occurred in chest region from radar signals~\cite{li2013review}. This section will briefly mention the principle of heart vibration measurement using different types of radars, with the analysis of the recent trends in radar usage.

\subsection{Types, Working Principles and Operating Frequencies}
Different radar systems transmit different types of waveforms as shown in Figure~\ref{fig:pipeline}(b). For example, continuous wave (CW) radar uses continuous wave with a fixed frequency, frequency modulated continuous wave (FMCW) radar uses continuous wave with linearly increased frequency, and impulse-radio ultra wide band (IR-UWB) radar uses pulses with wide frequency bandwidth. Readers are referred to references \cite{li2013review} and \cite{obadi2021survey} for the theoretical explanations of the radar principles, and to~\cite{obadi2021survey} and \cite{wang2020experimental} for the comparison between different radar types. The phase components of the transmitted signals for CW and FMCW radar are modulated by the displacement composed by respiration, heartbeat and all kinds of noises in a non-linear manner as proved in~\cite{chen2021movi,li2013review}. Then, the phase variation hidden in the raw received signal can be revealed using phase unwrapping techniques such as arctangent demodulation and extended differentiate and cross-multiply algorithm~\cite{obadi2021survey,wang2020remote}. For IR-UWB radar, the cardiac features are embedded in the propagation time delay of the echo signal~\cite{obadi2021survey}. 

Different radar types require different architectures and are suitable for different tasks. For example, CW radar has a simple architecture and adopts the fundamental baseband signal-processing methods, but the range information cannot be extracted from the received signals due to the lack of modulation~\cite{singh2020multi}. FMCW radar outperforms CW radar by leveraging the frequency modulation techniques, improving the signal-to-noise ratio (SNR) and providing the capability of range detection to further isolate the signal reflected only from the chest region~\cite{ha2020contactless}. Different from the continuous waveform used in CW and FMCW radar, IR-UWB radar emits widely spaced pulses with very short duration (e.g., $0.1-2$ ns)~\cite{hirt2003ultra}. Therefore, IR-UWB radar is more power-efficient than FMCW radar but normally cannot ensure a high SNR and range resolution~\cite{obadi2021survey}. In literature, IR-UWB radar is normally used for through-the-wall or long-distance monitoring~\cite{shen2018respiration}, but the complex radar architecture (e.g., requiring internal delay calibration~\cite{singh2020multi}) and signal-processing algorithms (e.g., harmonic rejection~\cite{wang2020experimental}) limit the relative research\cite{li2013review}.
        
In addition to the different radar types, radar operating frequency (carrier frequency) is another crucial parameter affecting cardiac monitoring quality because the frequency is inversely related to the beamwidth for antennas with the same diameter~\cite{huang2021antennas}, enabling the radar system with high operating frequency using narrow beamwidth to enhance directivity~\cite{ramasubramanian2018moving}.  Obeid \textit{et al.}~\cite{obeid2008low} found out that high operating frequency provides a large phase difference caused by heart vibration and improves the sensitivity of cardiac monitoring. The researchers in~\cite{ramasubramanian2018moving} claimed that the radar with a high operating frequency, especially the millimeter-wave (mmWave) range ($30-300$ GHz), can achieve a high range resolution, good noise-robustness and small antenna size.

\subsection{Trends in Radar Usage}\label{sec:trend}
\begin{figure}[tbp]
    \centering
    \subfloat[]{\label{fig:radar_type}\includegraphics[width=0.8\columnwidth]{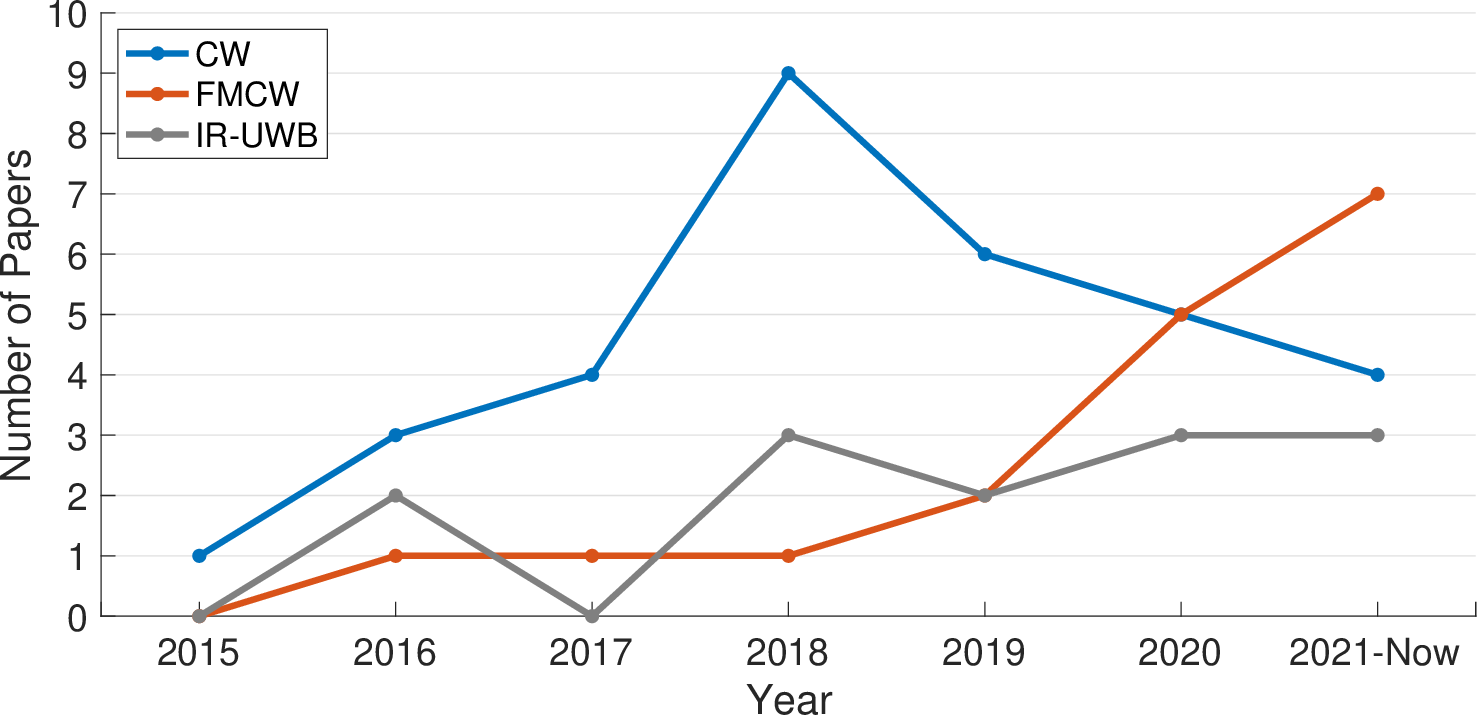}}\\
    \subfloat[]{\label{fig:radar_freq}\includegraphics[width=0.8\columnwidth]{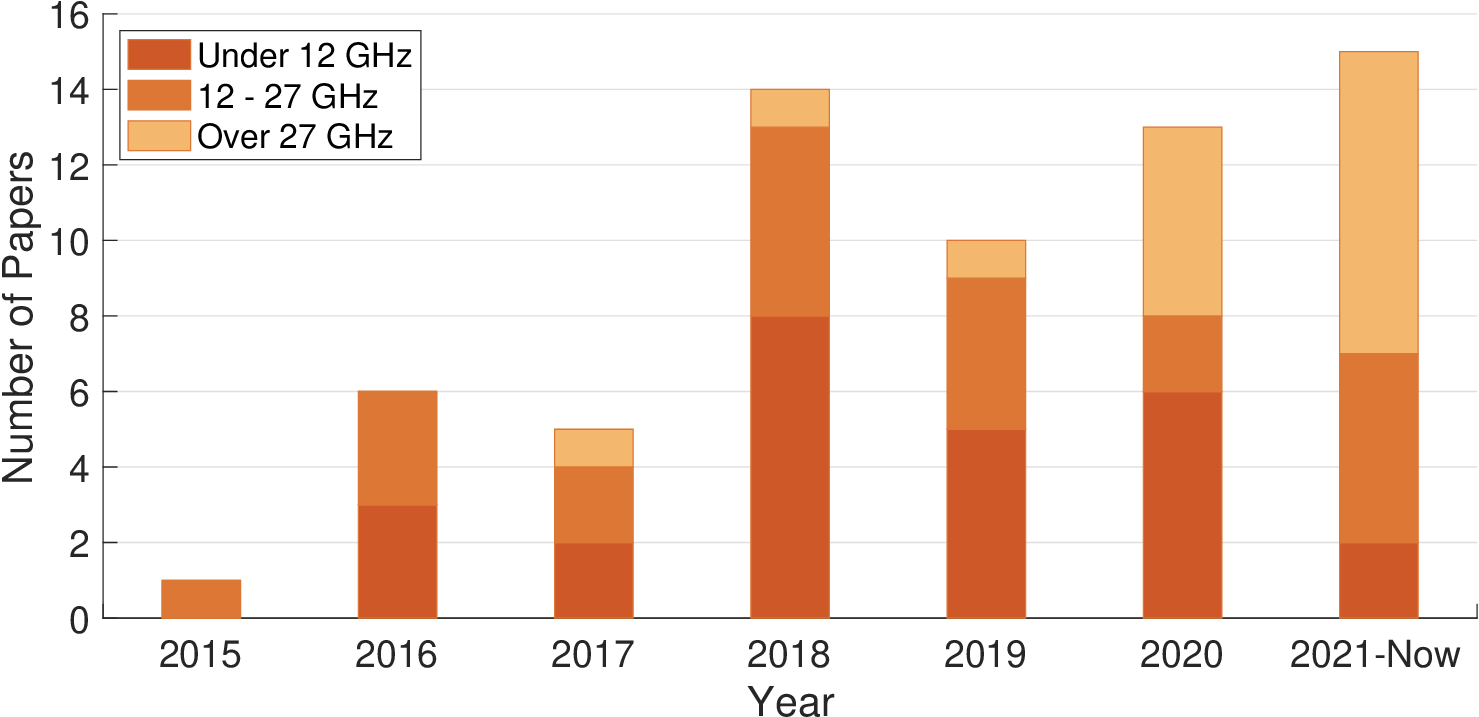}} 
    \caption{The trend of radar usage for cardiac monitoring after 2015: (a) the number of papers using different radar types; (b) the number of papers using different carrier frequencies.}
    \label{fig:freq_band_comp} 
\end{figure}

During the literature review, some valuable trends in radar usage are found and may help new researchers conduct their experiments. Figure~\ref{fig:radar_type} shows the statistics of the radar types adopted by the research mentioned in this review after 2015: CW radar was the most popular type before 2020 due to its simple architecture; FMCW radar receives a growing concern since 2015 because the recent-released commercial FMCW radar platforms reduce the knowledge required for designing or setting up a radar system~\cite{zhou2022towards}; IR-UWB radar is less popular than the other two types due to its complex radar architecture and signal-processing algorithms~\cite{singh2020multi,wang2020experimental}. For the trend in operating frequency selection, Figure~\ref{fig:radar_freq} shows that the early studies all focus on the low-frequency band for the simplicity of radar architecture design and baseband signal-processing algorithms, whereas the recent researchers are steering toward using mmWave-radar (especially $60$ or $77$ GHz commercial radar platform) for good performance. In summary, the trends revealed in Figure~\ref{fig:freq_band_comp} coincide with the evaluations from the last subsection, showing the FMCW with high operating frequency is becoming the mainstream for radar-based cardiac monitoring due to the balance between the performance and complexity.

\section{Cardiac Feature Extraction Algorithms}\label{sec:algos}
This section introduces the algorithms used in radar-based cardiac feature extraction following the structure of the newly proposed taxonomy, because the existing classification criteria are either too coarse to reveal the core features of the algorithms, or too narrow to cover the majority of the algorithms used for cardiac feature extraction. If the coarse categorizations based on the time/frequency domain or direct/indirect monitoring~\cite{mercuri2019vital} are adopted, numerous algorithms will be contained in each category, and researchers can hardly get any impression of these algorithms. In addition, some categorizations are restricted to specific applied scenarios. For example, Singh \textit{et al.}~\cite{singh2020multi} classified the algorithms based on the unsolved issues (e.g., RBM, respiration harmonics); Ye and Ohtsuki~\cite{ye2021spectral} separated the feature extraction process into three stages: frequency-domain transform, time-domain denoising and peak selection. The above mentioned scenario-driven categorizations in~\cite{ye2021spectral} and \cite{singh2020multi} only focus on certain applications, but fail to systematically cover most of the existing algorithms because some algorithms can be applied in different scenarios to solve multiple issues. In this review, the principles of all the algorithms with their variations and applications will be elaborated based on the proposed taxonomy, which classifies the algorithms with similar features or principles together to help the researchers and practitioners have a quick impression of each method, as shown in Table~\ref{tab:category}.

\begin{table*}[tb]
    \centering
    \caption{Classification of the algorithms introduced in this review}
      \begin{tabular}{l lll}
      \toprule
      \textbf{Category with Algorithm} & \multicolumn{3}{c}{\textbf{Reference$^1$ with Key Information$^{2,3}$}}\\
      \toprule
      \textbf{Spectrum-based Methods} &  \\
      \ \ 1. FFT with Improved Filter & \cite{petrovic2019high} ($2019$, $94$, CW, $24$ GHz) & \cite{saluja2019supervised} ($2019$, $48$, CW, $5.8$ GHz) &\cite{yang2020vital} ($2020$, $20$, CW, $10$ GHz)\\ & \cite{zhu2018fundamental} ($2018$, $36$, CW, $2.4$ GHz) \\
      \ \ 2. FFT with Differentiator & \cite{le2020heartbeat} ($2020$, $6$, IR-UWB, $4.6$ GHz) & \cite{zhao2016emotion} ($2016$, $444$, FMCW, $5.5$ GHz) &\cite{xiong2020differential} ($2020$, $35$, FMCW, $7.9$ GHz) \\
      \ \ 3. Short-time Fourier Transform & \cite{hu2014real} ($2015$, $11$, CW, $5.8$ GHz)& \cite{xiong2017accurate} ($2017$, $57$, CW, $10.5$ GHz)& \cite{tu2015fast} ($2015$, $104$, CW, $5.8$ GH)\\
      \ \ 4. Discrete Cosine Transform &  \cite{park2017polyphase} ($2019$, $42$, CW, $10.2$ GHz)& \cite{shih2021quadrature} ($2020$, $5$, CW, $2.4$ GHz)& \\
      \ \ 5. Wavelet Transform & \cite{tomii2015heartbeat} ($2015$, $36$, CW, $24$ GHz) &\cite{li2017wavelet} ($2017$, $91$, CW, $5.8$ GHz) &\cite{mercuri2019vital} ($2019$, $154$, FMCW, $5.8$ GHz)\\ & \cite{wang2020remote} ($2020$, $82$, FMCW, $77$ GHz)& \cite{liu2022vital} ($2022$, $4$, IR-UWB, $39$ GHz) & \cite{ling2022non} ($2022$, $2$, FMCW, $77$ GHz)\\
      \toprule
      \textbf{Periodicity-based Methods} &  \\
      \ \ 1. Derivative-based Peak Detection & \cite{kim2019peak} ($2019$, $42$, CW, $77$ GHz)& \cite{zhang2020health} ($2020$, $4$, CW, $60$ GHz) &\cite{mogi2017heartbeat} ($2017$, $37$, CW, $24$ GHz) \\ & \cite{yamamoto2018spectrogram} ($2018$, $32$, CW, $24$ GHz) & \cite{xu2021accurate} ($2022$, $13$, IR-UWB, $7.3$ GHz) &\cite{ye2021spectral}\;\;\  ($2021$, $6$, CW, $24$ GHz)  \\ &\cite{will2018radar} ($2018$, $90$, CW, $24$ GHz)\\      
      \ \ 2. Auto-correlation & \cite{wang2020remote} ($2020$, $82$, FMCW, $77$ GHz)& \cite{nosrati2017high} ($2017$, $89$, CW, $2.4$ GHz)& \cite{ha2021wistress} ($2021$, $11$, FMCW, $77$ GHz) \\
      \ \ 3. Cross-correlation with Template & \cite{will2016instantaneous} ($2016$, $13$, CW, $24$ GHz)& \cite{lv2018doppler} ($2018$, $68$, CW, $5.8$ and $24$ GHz)& \cite{ha2020contactless} ($2020$, $46$, FMCW, $77$ GHz) \\& \cite{zhang2020health} ($2020$, $4$, CW, $60$ GHz)& \cite{zhao2016emotion} ($2016$, $444$, FMCW, $5.5$ GHz)& \cite{chen2021contactless}\;\;\ ($2022$, $7$, FMCW, $77$ GHz) \\&  \cite{will2017advanced} ($2017$, $32$, CW, $24$ GHz) &\cite{sakamoto2015feature} ($2015$, $97$, IR-UWB, $26.4$ GHz) \\
      \ \ 4. Hidden Markov Model & \cite{xia2021radar} ($2021$, $19$, CW, $24$ GHz)&\cite{mei2020fast} ($2020$, $2$, FMCW, $2.4$ GHz)& \cite{will2018radar} ($2018$, $90$, CW, $24$ GHz) \\& \cite{shi2021contactless} ($2021$, $18$, CW, $24$ GHz)  \\
      \toprule
      \textbf{Blind Source Separation Methods} &  \\
      \ \ 1. Multiple Signal Classification & \cite{lee2016tracking} ($2016$, $36$, CW, $24$ GHz)& \cite{bechet2013non} ($2015$, $28$, CW, $2.4$ GHz)& \cite{yamamoto2018non} ($2018$, $7$, CW, $24$ GHz) \\& \cite{yamamoto2019music} ($2019$, $6$, CW, $24$ GHz) \\
      \ \ 2. Independent Component Analysis & \cite{mercuri2018direct} ($2018$, $60$, CW, $5.8$ GHz)& \cite{mercuri2021enabling} ($2021$, $21$, FMCW, $7.3$ GHz)& \cite{xu2021accurate} ($2022$, $13$, IR-UWB $7.3$ GHz) \\& \cite{lv2021non} ($2021$, $19$, FMCW, $120$ GHz) \\
      \ \ 3. Empirical Mode Decomposition  & \cite{liu2020vital} ($2020$, $10$, FMCW, $77$ GHz)& \cite{diraco2017radar} ($2017$, $97$, IR-UWB, $4.3$ GHz) & \cite{zhang2020health} ($2020$, $4$, CW, $60$ GHz) \\& \cite{sun2020remote} ($2020$, $31$, FMCW, $77$ GHz)& \cite{shyu2018detection} ($2018$, $76$, IR-UWB, $4.3$ GHz) & \cite{shyu2020uwb} ($2020$, $12$, IR-UWB, $4.3$ GHz) \\
      \ \ 4. Variational Mode Decomposition & \cite{shen2018respiration}\;\;\  ($2018$, $77$, IR-UWB, $4.3$ GHz)& \cite{duan2018non} ($2018$, $61$, IR-UWB, $2.9$ GHz)& \cite{zhang2021mutual} ($2021$, $21$, FMCW, $77$ GHz)\\ & \cite{wang2021mmhrv} ($2021$, $24$, FMCW, $77$ GHz) \\
      \ \ 5. Sparse Signal Reconstruction &  \cite{zhang2021mutual} ($2021$, $21$, FMCW, $77$ GHz) & \cite{wang2019noncontact} ($2021$, $24$, IR-UWB, $5$ GHz) &\cite{ye2018robust} ($2018$, $8$, CW, $24$ GHz)  \\&  \cite{ye2018stochastic} ($2018$, $30$, CW, $24$ GHz) & \cite{ye2019blind} ($2019$, $33$, CW, $24$ GHz) \\
      \toprule
      \textbf{Deep Learning Methods} &  \\
      \ \ 1. Convolutional Neural Network & \cite{ha2020contactless} ($2020$, $46$, FMCW, $77$ GHz)& \cite{chen2021movi} ($2021$, $30$, FMCW, $77$ GHz) & \cite{chen2021contactless}\;\;\ ($2022$, $7$, FMCW, $77$ GHz)\\& \cite{wu2019person} ($2019$, $27$, IR-UWB, $79$ GHz) \\
      \ \ 2. Long Short-term Memory & \cite{yamamoto2020ecg} ($2022$, $7$, CW $24$ GHz) &\cite{gong2021rf} ($2021$, $6$, FMCW, $77$ GHz)& \cite{shi2021contactless} ($2019$, $10$, CW, $24$ GHz)\\& \cite{li2019standalone} ($2019$, $10$, CW, $5.8$ GHz) &\cite{ye2021spectral}\;\;\ ($2021$, $6$, CW, $24$ GHz) \\
      \ \ 3. Deep Contrastive Learning & \cite{chen2021movi} ($2021$, $30$, FMCW, $77$ GHz) \\
      \bottomrule
      \multicolumn{4}{l}{1. The same paper might occur multiple times in this table if multiple algorithms are adopted.} \\
      \multicolumn{4}{l}{2. The key information includes the year of publication, times cited (before April, 2023), radar type and operating frequency.} \\
      \multicolumn{4}{l}{3. The operating frequency refers to the starting frequency for FMCW radar or the central frequency for IR-UWB radar.}
      \end{tabular}%
    \label{tab:category}%
\end{table*}

\subsection{Spectrum-Based Methods}
\begin{figure}[tbp]
    \centering
    \subfloat[]{\label{fig:rawRadar}\includegraphics[width=0.45\columnwidth]{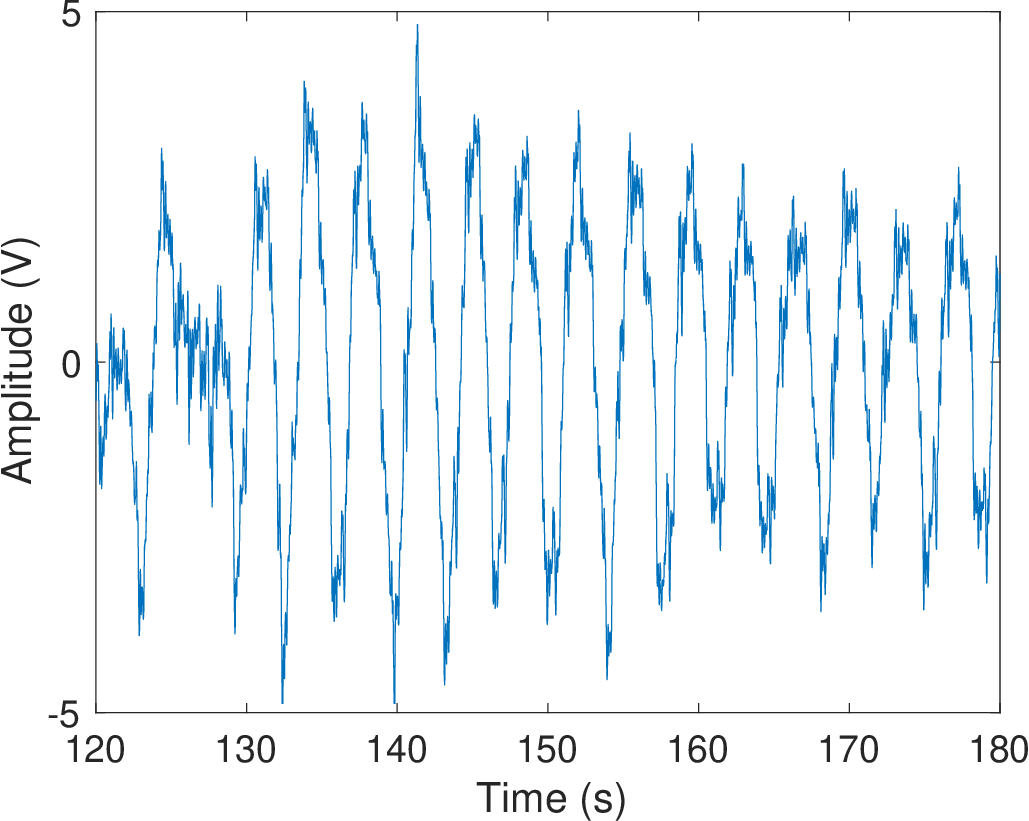}}
    \subfloat[]{\label{fig:rawRadar_fft}\includegraphics[width=0.45\columnwidth]{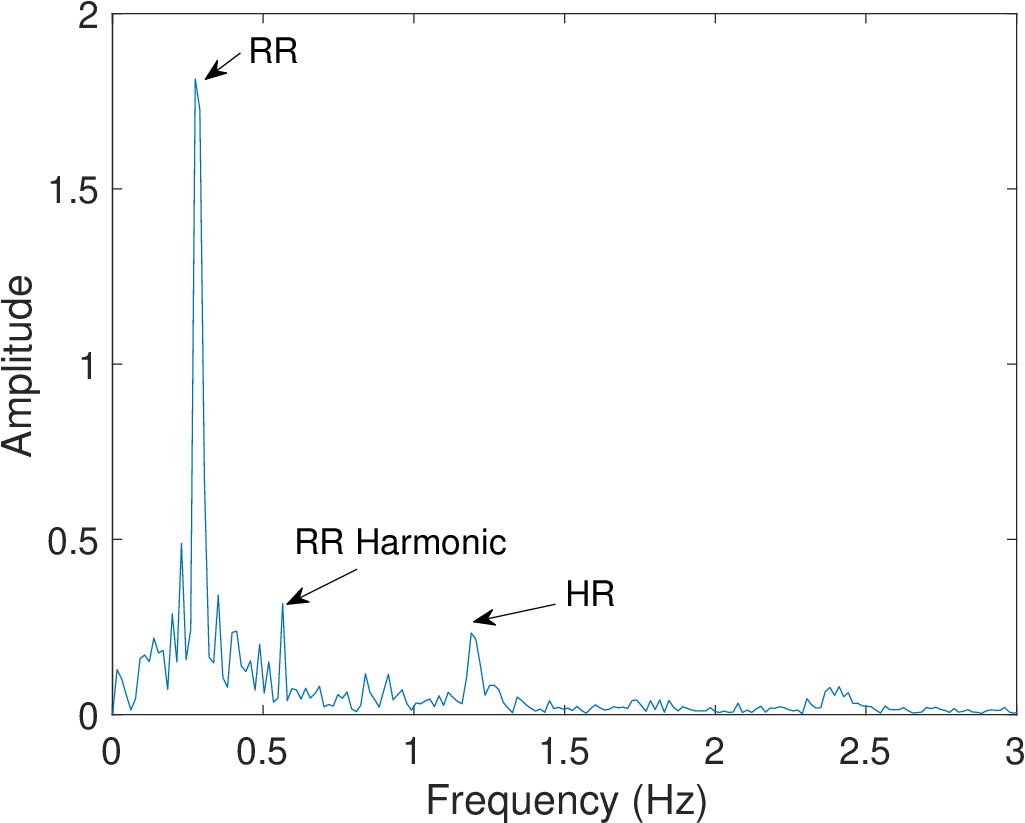}} \\ 
    \subfloat[]{\label{fig:rawRadar_diff}\includegraphics[width=0.45\columnwidth]{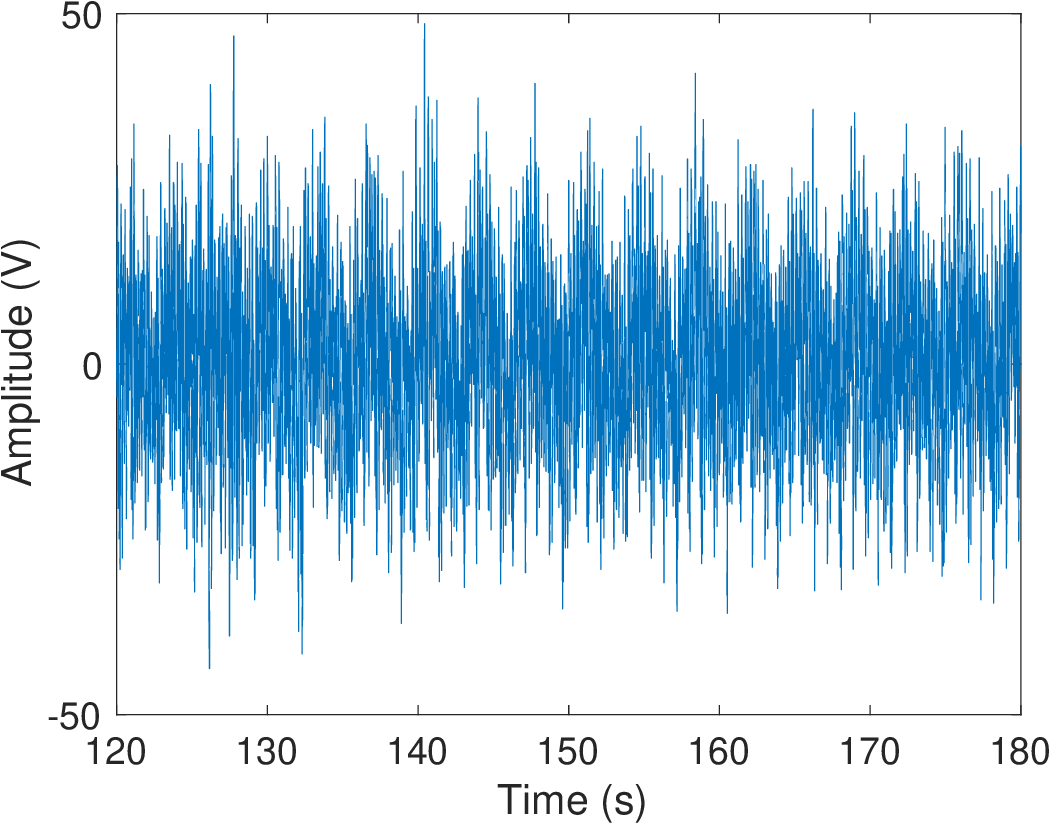}}
    \subfloat[]{\label{fig:rawRadar_diff_fft}\includegraphics[width=0.45\columnwidth]{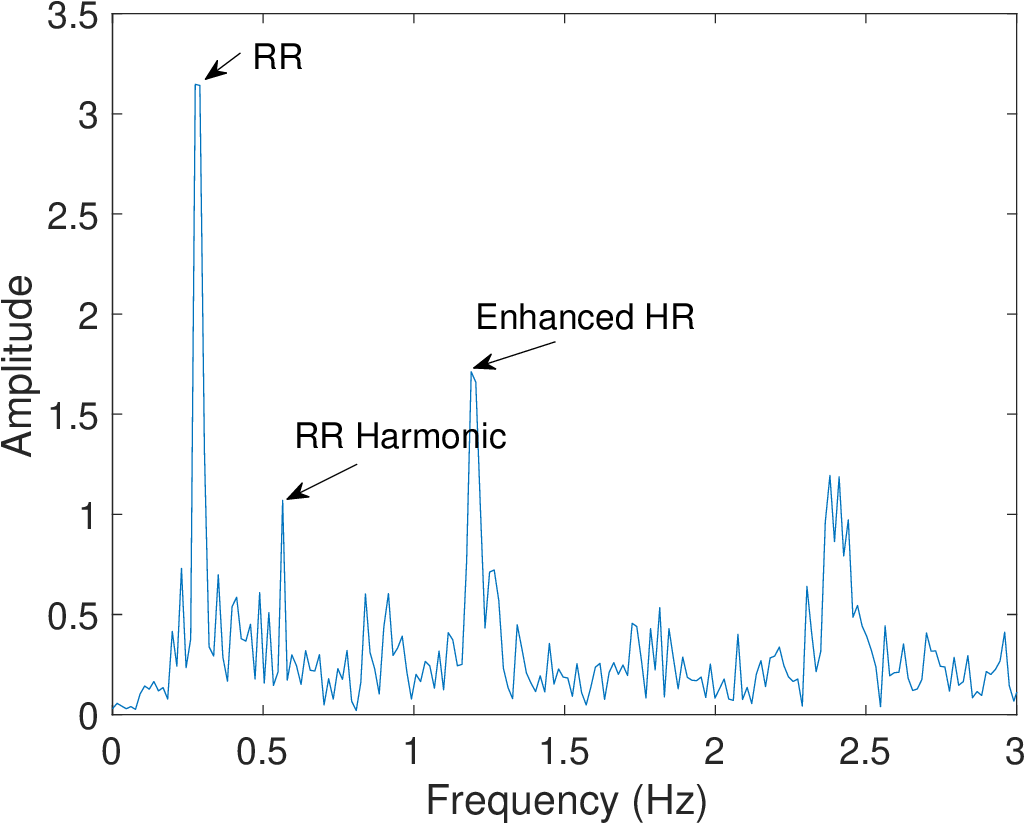}} \\ 
    \caption{Illustration for spectrum-based methods: (a) raw radar signal; (b) spectrum obtained for raw radar signal from FFT, with RR, RR harmonic and HR peaks labelled; (c) raw radar signal after differentiation; (d) spectrum obtained for differentiated raw radar signal from FFT, with enhanced HR peak labelled.}
    \label{fig:fftForHR}
\end{figure}

Spectrum represents the transformation of signal from time-domain to frequency-domain and can reveal the main frequency components of the signal. Normally, the respiration rate (RR) and HR frequency at rest are in the range of $0.1-0.5$ Hz and $1-1.6$ Hz respectively~\cite{mogi2017heartbeat}, making the spectrum-based method the most straightforward way to extract cardiac features within a certain frequency range. The spectrum-based methods usually share similar ideas: (a) a piece of unwrapped phase signal taken from the dataset only shows the periodic displacement induced by strong respiration instead of subtle cardiac activities as shown in Figure~\ref{fig:rawRadar}; (b) the raw signal is then transformed into a spectrum using time-frequency transformation methods (e.g., FFT) as shown in Figure~\ref{fig:rawRadar_fft} with corresponding peaks labelled; (c) for the low-noise scenario, after filtering the frequency components of RR, RR harmonic and other high-frequency noise, the remaining dominant peak of the spectrum represents the HR frequency; (d) for the noisy scenario, the researchers could use several techniques, such as noise-cancellation filter or differentiator, to suppress the noise or enhance the HR frequency component on the spectrum.

\subsubsection{Fast Fourier Transform with Improved Filter}\label{sec:filter}
Conventional filter with fixed passband is unsuitable for estimating the HR frequency varying within a wide range~\cite{gong2021rf} or filtering the noises with the frequency components falling in the passband of HR, such as high-order RR harmonics~\cite{xiong2017accurate} and engine vibration~\cite{zheng2020v2ifi}. For the improvements of filters, Petrovic \textit{et al.}~\cite{petrovic2019high} designed a BPF bank with different passbands and selected the optimal BPF used for each estimation in terms of the coarse HR estimated by chirp Z-transform~\cite{rabiner1969chirp}. Saluja \textit{et al.}~\cite{saluja2019supervised} considered the natural intrinsic feature of the RR system and trained a gamma-filter using the ground-truth ECG measurements to remove the RR components. In addition, to realize the undistorted noise reduction, the adaptive-noise-cancellation filter~\cite{widrow1975adaptive} can be used with the aid of prior-knowledge about noises. In~\cite{yang2020vital}, the proposed adaptive-noise-cancellation filter can mitigate RBM noise for new inputs after being trained with the noise synthesized by the method of polynomial fitting and optimized by the least-mean-square algorithm. Similarly, Zhu \textit{et al.}~\cite{zhu2018fundamental} trained the adaptive-noise-cancellation filter with the radar movement noise measured by a dual-frequency radar and realized the cardiac monitoring on a moving platform.

\subsubsection{Fast Fourier Transform with Differentiator}
In real-world applications, the weak heart vibration can be easily drowned out by respiration noise with orders of magnitude larger than the heartbeat vibration, whereas the acceleration of the displacement induced by the heartbeat is larger than that of the respiration~\cite{zhao2016emotion}. Therefore, to simultaneously suppress the RR harmonics and enhance the HR components on the spectrum obtained from FFT, differentiator can be applied to extract the acceleration information as shown in Figure~\ref{fig:rawRadar_diff} and~\ref{fig:rawRadar_diff_fft}. The performance of the differentiators with different orders ($0^{th}$ to $3^{rd}$ order) was compared in~\cite{le2020heartbeat}, with the result illustrating that third-order differentiator could achieve the best SNR and hence the HR estimation. However, Ren \textit{et al.}~\cite{ren2021vital} claimed that the high-order differentiator may also amplify the high-frequency noise and further proposed the first and second order noise-robust differentiator by calculating the derivatives using six adjacent samples with corresponding time intervals, instead of using only two samples. Zhao \textit{et al.}~\cite{zhao2016emotion} applied the second-order differentiator in a less-noisy scenario to suppress the RR harmonics and revealed the periodic pattern for each heartbeat cycle. However, to avoid enhancing high-frequency noise for low-SNR scenarios, Xiong \textit{et al.}~\cite{xiong2020differential} first assessed the signal SNR according to the sparseness of the amplitude spectrum obtained from FFT and then adaptively switched between the first and second-order differentiators before the HR estimation. 

\subsubsection{Short-Time Fourier Transform}
FFT is a conventional method to obtain the spectrum of signal, whereas the high-quality spectrum for HR estimation requires: (a) appropriate truncation of the signal to avoid spectrum leakage and smearing issue~\cite{ren2016phase}; (b) long observation period (at least 15 sec) to ensure the spectrum with high frequency resolution~\cite{tu2015fast}. To circumvent the above mentioned two requirements, short-time Fourier transform (STFT) has been proven capable of avoiding spectrum leakage and providing time-related spectrum using short window length~\cite{mercuri2013analysis}. In~\cite{hu2014real}, by adding a twelve-second sliding Hamming window on the original signal with further interpolation, the high frequency resolution ($0.05$ Hz) is achieved with a limited data length. Similarly, Xiong \textit{et al.}~\cite{xiong2017accurate} performed STFT and realized the HR monitoring using an eight-second sliding window under minor RR harmonics. To further shorten the window length and resist RR harmonics, Tu \textit{et al.}~\cite{tu2015fast} found that the HR signal is immune to the time-window variation compared with the RR signal and proposed a time-window-variation technique to estimate the HR according to the spectra obtained from multiple adjacent time windows (from $2$ to $5$ sec).

\subsubsection{Discrete Cosine Transform}
The short window length used in STFT could improve the real-time monitoring performance but may increase the main-lobe width and side-lobe width~\cite{park2017polyphase}, causing the overlapping of the lobes on the spectrum. To narrow the main-lobe width and side-lobe width, discrete cosine transform (DCT)~\cite{ahmed1974discrete} evolving from discrete Fourier transform is proposed to increase the estimation accuracy. Park \textit{et al.}~\cite{park2017polyphase} claimed that the misaligned phase of the input signal affects the HR estimation and proposed polyphase-basis-DCT (PB-DCT) by simultaneously calculating the estimated HR using eight bases with different phases, realizing the HR estimation using a $1.5$ sec time window. Furthermore, to circumvent the simultaneous calculations in PB-DCT and solve the null point issue~\cite{droitcour2004range}, Shih \textit{et al.}~\cite{shih2021quadrature} designed the quadrature cosine transform with varying window length alignment technique to avoid the degradation caused by misaligned phase, and hence reduce the non-linear interference.

\subsubsection{Wavelet Transform}
\begin{figure}[tb] 
    \centering 
    \includegraphics[height=0.6\columnwidth, width=0.7\columnwidth]{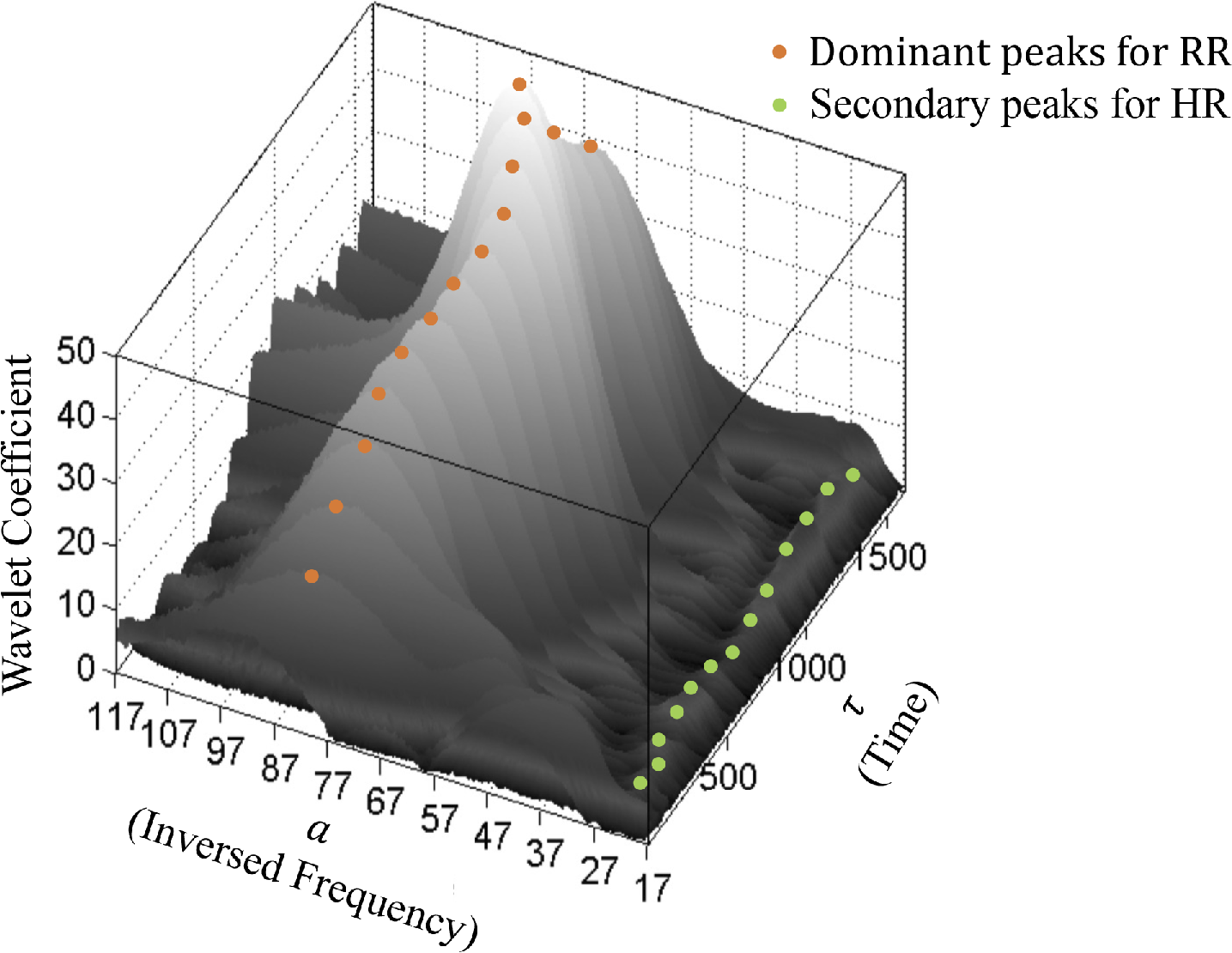} 
    \caption{Illustration for WT with the dominant peaks for RR and secondary peaks for HR~\cite{tariq2011vital}.}
    \label{fig:wt_pin}  
\end{figure}

Wavelet transform (WT) is a famous time-frequency analysis tool~\cite{bentley1994wavelet} that outperforms STFT in terms of transient signals analysis, because WT adopts the adjustable mother wavelet as the basis to calculate the wavelet coefficient and has two adjustable parameters $a$ and $\tau$ acting as scale and shift factors respectively. By varying $a$ (to dilate or compress the signal) and $\tau$ (to shift the wavelet along the time axis), the mixed signal can be decomposed simultaneously in both frequency and time domain as shown in Figure~\ref{fig:wt_pin}, with the small peak ($a\approx (17,27)$) representing the HR variation along time~\cite{tariq2011vital}.

In literature, WT can be utilized as the main algorithm for cardiac feature extraction. Tomii \textit{et al.}~\cite{tomii2015heartbeat} took the power of the received signal as the input of WT and selected the scale factor producing the most peaks on the spectrum as the optimal scale factor to mitigate the RBM noise. Li \textit{et al.}~\cite{li2017wavelet} first compared the performances of different mother wavelets and selected Morlet wavelet. Then, to reveal the HR peak ruined by respiration, the RR harmonics were identified by simultaneously performing WT on the signals with different lengths because the peaks of RR harmonics would shift on the spectrum while the peaks of HR would not. Zhao \textit{et al.}~\cite{zhao2017noncontact} proposed synchrosqueezing transform by reassigning the energy distribution of WT to solve the smearing issue brought by the wavelet function and also improve the frequency resolution. 

WT can also assist with other methods to realize pre-denoising. In~\cite{mercuri2019vital}, the artifacts were located on the time-frequency plot obtained by WT and then attenuated by the moving average filter. Wang \textit{et al.}~\cite{wang2020remote} first separated RR and HR signals based on the scale factor and then applied the rigrsure adaptive soft threshold for noise reduction, reserving a good continuity at the threshold boundary. Similarly, Liu \textit{et al.}~\cite{liu2022vital} used the same soft threshold to cancel the noise on the time scale, but the hard threshold was further added to denoise the frequency scale and maintain the peak characteristics~\cite{liu2022vital}. To remove the noise from the non-stationary signals, Ling \textit{et al.}~\cite{ling2022non} applied empirical wavelet transform to imitate the BPF with multiple sub-passbands using wavelet-based filter banks.

\subsection{Periodicity-Based Methods}
Based on the natural periodicity of the cardiac features, the methods introduced in this subsection leverage probability model~\cite{xia2021radar} or template determined by cardiac morphology~\cite{will2017advanced} to identify the periodical patterns obscured by noise in radar signals~\cite{sakamoto2015feature}.

\subsubsection{Derivative-Based Peak Detection}
\begin{figure}[tb] 
    \centering 
    \includegraphics[height=0.5\columnwidth]{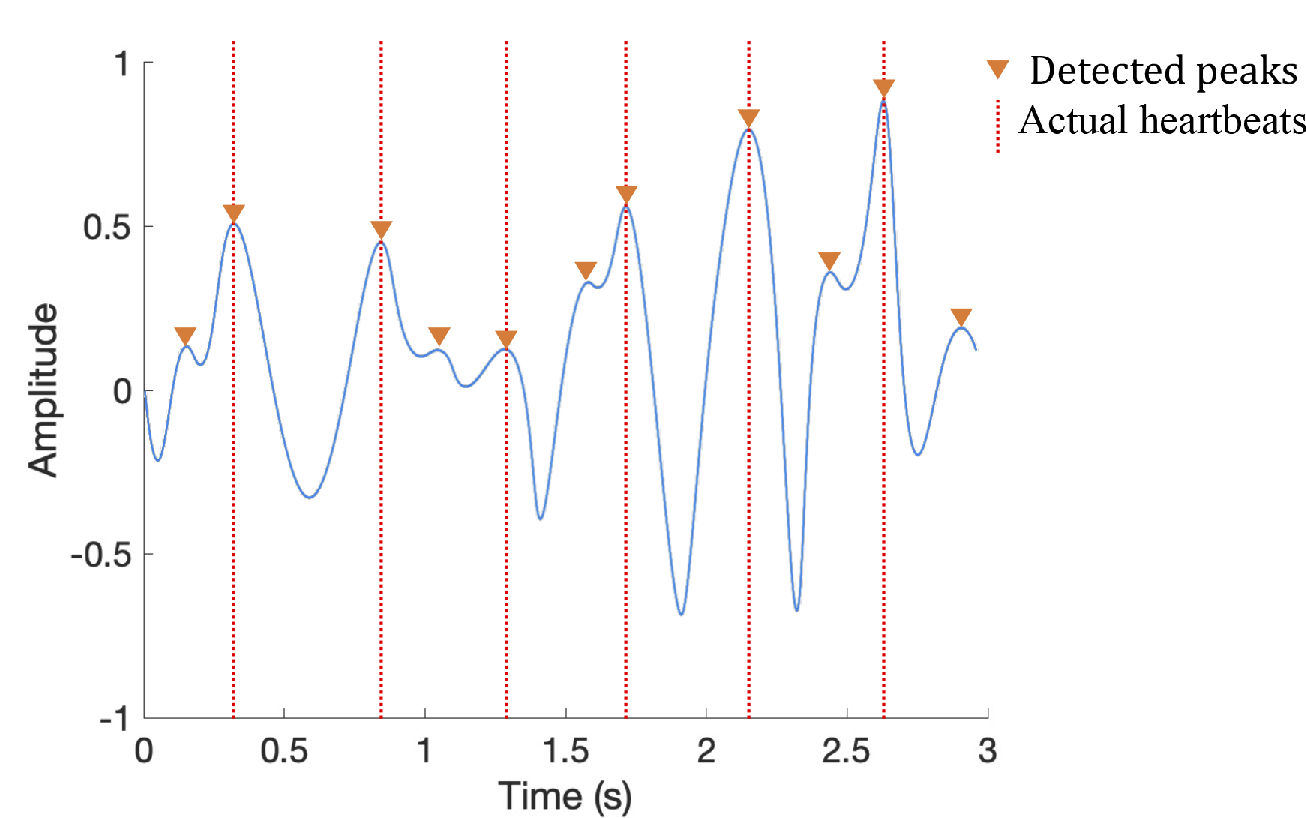} 
    \caption{Illustration for peak detection with orange triangles for all detected peaks and red lines for heartbeats.}
    \label{fig:peak_dete} 
\end{figure}

The heartbeats will generate periodically occurred peaks on time- or frequency-domain signals, while the peaks for the noises appear randomly. Although both types of peaks (indicated by orange triangles in Figure~\ref{fig:peak_dete}) can be detected through derivative calculation, additional peak detection algorithms are still required to distinguish the peaks of heartbeats (indicated by red lines in Figure~\ref{fig:peak_dete}) from others.

According to~\cite{mei2020fast}, the measurement of BBI is almost the same for the adjacent heartbeats and is a crucial feature to select the right peaks. Therefore, based on the empirical BBI, Kim \textit{et al.}~\cite{kim2019peak} selected the peaks in the time-domain signal by setting an amplitude threshold to identify the potential peak candidates and a temporal threshold to filter the peaks that occur too closely. Zhang \textit{et al.}~\cite{zhang2020health} adopted a similar peak-selection strategy, but the time-domain signal was pre-filtered with an adaptive passband determined by the dominant peak on the spectrum to increase SNR. To mitigate the RR and RBM noise, Mogi \textit{et al.}~\cite{mogi2017heartbeat} set the thresholds based on empirical BBI, but the time-domain signal was converted from the spectrum integrated from $8$ to $50$ Hz. Yamamoto \textit{et al.}~\cite{yamamoto2018spectrogram} first determined the time window that contains around three heartbeats based on the previous BBI, and then each BBI was determined by three detected peaks, ensuring a stable HR estimation.

In addition, some peak detection methods are implemented on the frequency spectrum. Xu \textit{et al.}~\cite{xu2021accurate} thought that the high-order harmonics of the heartbeat also contain vital information and could naturally avoid RR harmonics. Therefore, the peaks for fundamental frequency and first/second harmonics were chosen on the spectrum within the range of $1.7$-$6.7$ Hz to produce several HR estimations with confidence scores. Ye \textit{et al.}~\cite{ye2021spectral} modelled the peak selection problem as dynamic programming. Multiple candidate paths were first created according to the peaks on the spectrum, and the optimal path was then selected using Viterbi algorithm based on the cost evaluated by branch/path metric. Will \textit{et al.}~\cite{will2018radar} proposed decoding peak detection by modelling the peak detection task as a probability estimation problem. The decoding peak detection method first defined two hidden states for the peaks on the frequency envelogram, with the probability of each observed state calculated by a logistic function to enhance the difference between states. Then, the most likely sequential states were decoded by the modified Viterbi algorithm and used for BBI calculation.

\subsubsection{Auto-correlation}
\begin{figure}[tbp]
    \centering
    \subfloat[]{\label{fig:rawRadar_auto}\includegraphics[width=0.35\columnwidth]{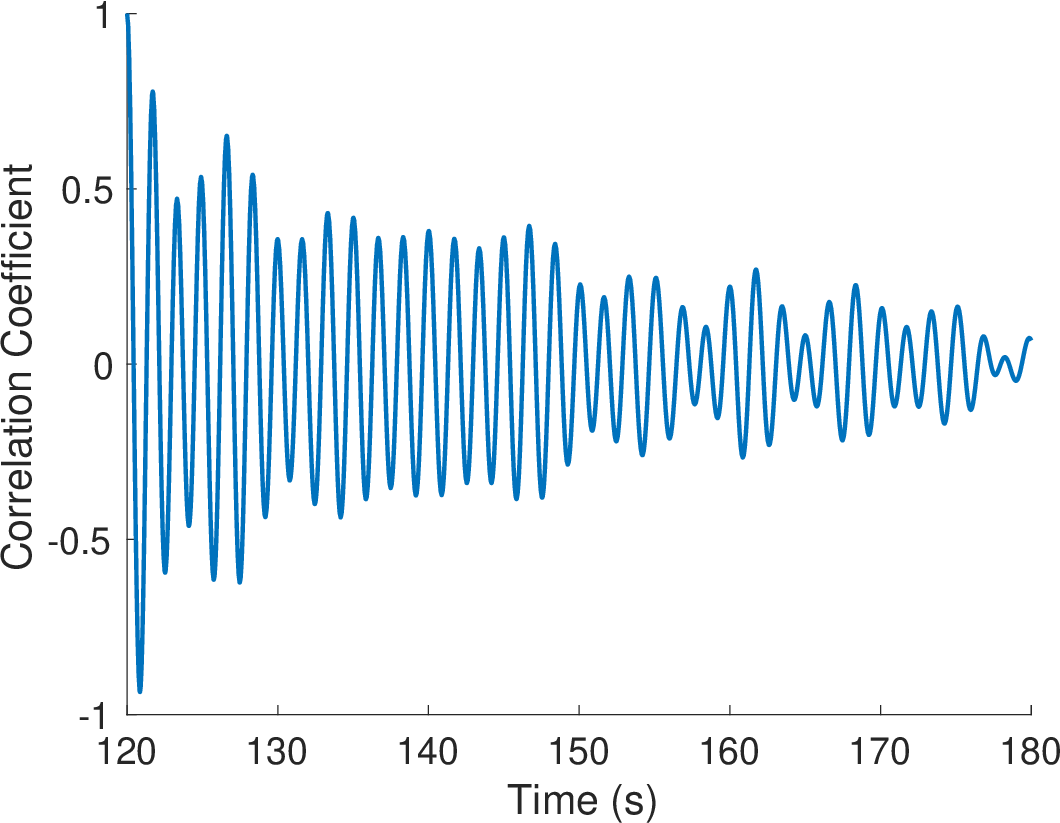}}
    \subfloat[]{\label{fig:rawRadar_auto_fft}\includegraphics[width=0.35\columnwidth]{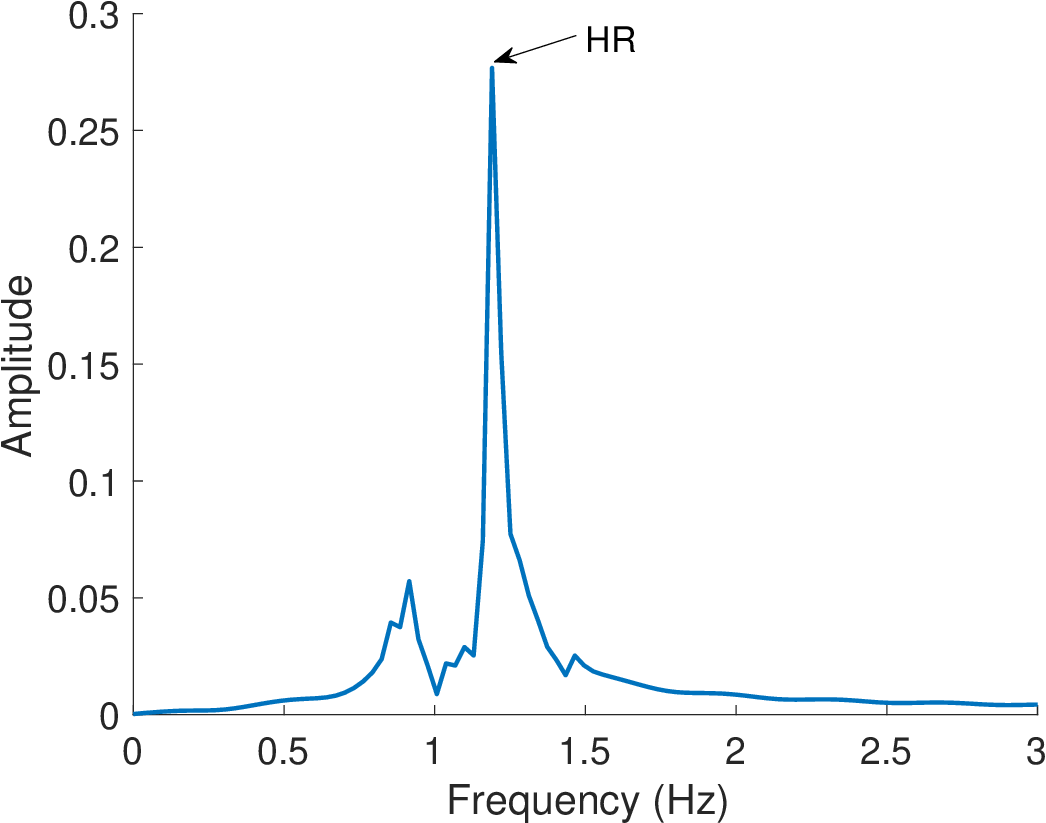}} 
    \subfloat[]{\label{fig:sm_fig}\includegraphics[width=0.3\columnwidth]{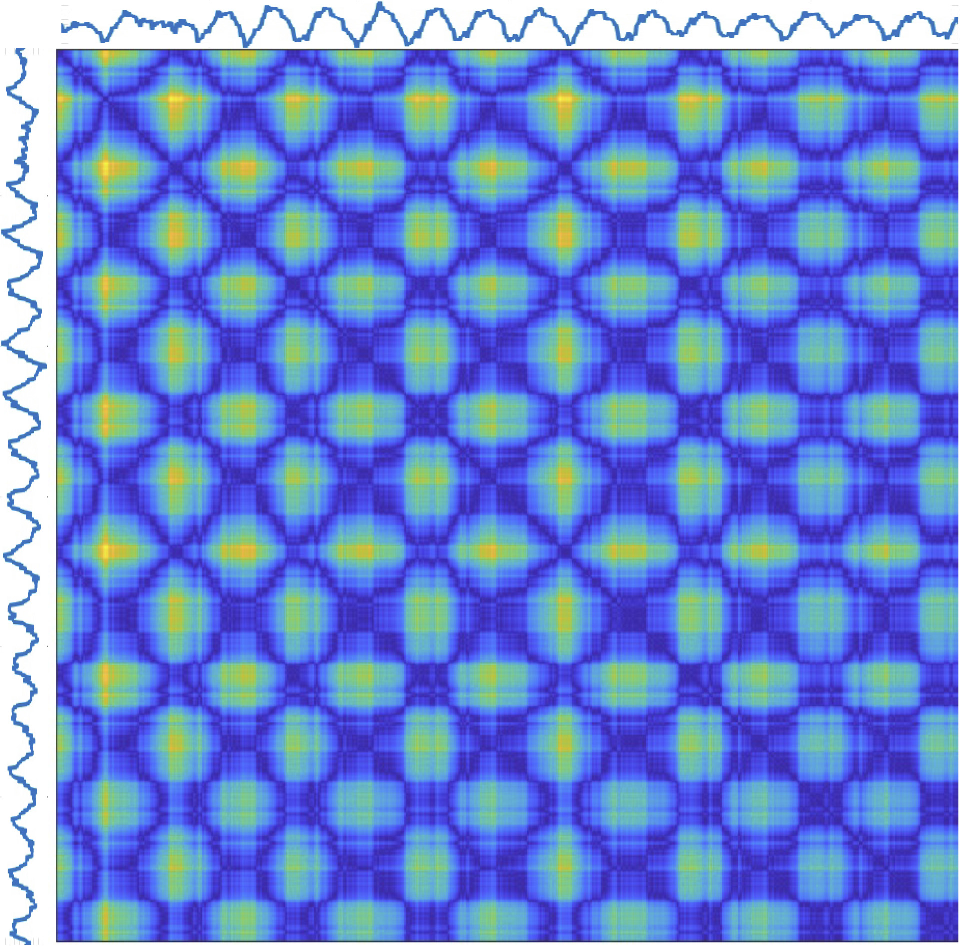}}
    \caption{Illustration for auto-correlation methods: (a) auto-correlation of filtered signal; (b) HR peak estimated by auto-correlation method; (c) heatmap for the SMM calculated from two identical raw radar signals, showing the repeated patterns for the heartbeats.}
    \label{fig:fftForACF}
\end{figure}

Due to the periodicity of the cardiac events, the time-domain heartbeat signal can be simplified as a periodic sinusoidal function~\cite{shen2018respiration} and recovered by an auto-correlation function (ACF) which aims to find similar patterns from the input time-domain signals. To illustrate, after filtering the RR components, the ACF of the filtered signal goes to zero with the dominant frequency representing the HR frequency, as shown in Figure~\ref{fig:rawRadar_auto} and~\ref{fig:rawRadar_auto_fft}. 

For the relative applications of auto-correlation method, Wang \textit{et al.}~\cite{wang2020remote} first pre-processed the raw signal by WT to enhance the periodicity and then applied ACF to the resultant signal to estimate the HR based on the peak intervals. To increase the frequency resolution, Nosrati \textit{et al.}~\cite{nosrati2017high} calculated the ACF of the complex time-domain unwrapped phase signal to reveal the periodicity and estimated the HR frequency with frequency-time phase regression technique. Ha \textit{et al.}~\cite{ha2021wistress} borrowed the idea from ACF and proposed a novel method to plot the heatmap according to the self-similarity matrix calculated with the two copies of the phase signal, as shown in Figure~\ref{fig:sm_fig}. The heatmap shows repeatedly appearing patterns for the heartbeats and can be processed by 2D convolutional neural network (CNN) to estimate the BBI. However, in most research, the auto-correlation method is not applied as the core algorithm for cardiac feature extraction, because the input signal is normally longer than ten seconds to reveal periodicity, and the repeated features are easily corrupted by noises~\cite{ha2021wistress}.

\subsubsection{Cross-correlation with Template}
\begin{figure}[tbp]
    \centering
    \subfloat[]{\label{fig:temp_time}\includegraphics[width=0.25\columnwidth]{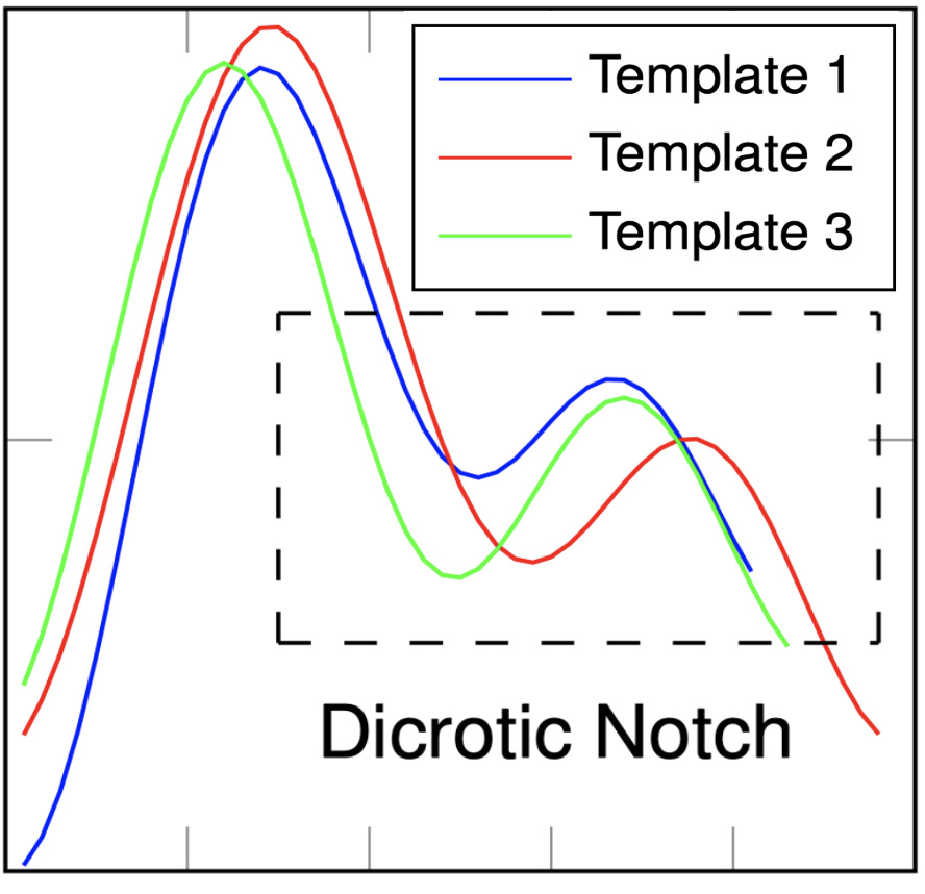}}
    \subfloat[]{\label{fig:temp_fea}\includegraphics[width=0.7\columnwidth]{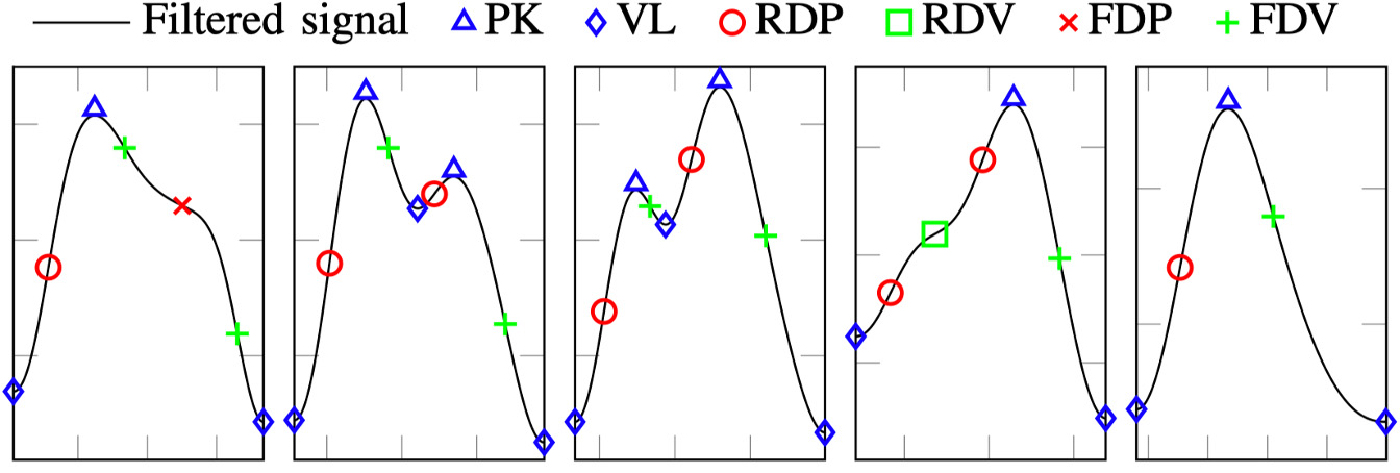}} 
    \caption{Two types of templates: (a) time-domain signals with prominent waveform~\cite{will2016instantaneous}; (b) sequential features extracted from signal~\cite{will2017advanced}.}
    \label{fig:template}
\end{figure}

Instead of finding the repeated patterns from time-domain signal with ACF, the methods in this part aim to identify the heartbeats by measuring the similarity between the pre-defined template based on cardiac morphology and the signal segments using the cross-correlation function (CCF). The intuition behind template-based methods is that the morphology for the consecutive human heartbeats is normally the same, only with certain stretch or compression~\cite{zhao2016emotion}. Therefore, it is possible to locate each heartbeat using templates formed either by time-domain signals with certain prominent waveform~\cite{will2016instantaneous} as shown in Figure~\ref{fig:temp_time} or by the features~\cite{will2017advanced} extracted from the time-domain waveform as shown in Figure~\ref{fig:temp_fea}.  

For the first type, Will \textit{et al.}~\cite{will2016instantaneous} selected the dicrotic notch as the template as shown in Figure~\ref{fig:temp_time} and applied CCF to measure the similarity between the template and the segments from the filtered radar signal. To mitigate large-scale RBM, Lv \textit{et al.}~\cite{lv2018doppler} captured the template during the static state and applied the template cleaned by polynomial fitting to locate the heartbeat for good SNR. Ha \textit{et al.}~\cite{ha2020contactless} proposed a CNN-assisted template matching with the kernel of CNN acting as the template, and the CNN model needs to be trained to get the optimal template (kernel) for the training dataset. However, the fixed template cannot detect abnormal cardiac events (e.g., heart arrhythmia) because the template is determined only for normal cardiac signals. Therefore, the adaptive template was proposed~\cite{zhang2020health,zhao2016emotion,chen2021contactless} according to the previously evaluated segments, with the linear time wrapping technique~\cite{zhao2016emotion} used to ensure the same length for different iterations. 

For the second type, Will \textit{et al.}~\cite{will2017advanced} first calculated the derivative of the signals and defined six features as peak, valley, rising derivative peak/valley and falling derivative peak/valley, as shown in Figure~\ref{fig:temp_fea}. Then, the CCF calculation is implemented based on the template formed by these six sequential features instead of the amplitude of the time-domain signals. Sakamoto \textit{et al.}~\cite{sakamoto2015feature} also adopted the same six features, whereas a complex value (phase) was assigned for each feature point to eliminate the unreliable feature point.

\subsubsection{Hidden Markov Model}\label{sec:hmm}
\begin{figure}[tb] 
    \centering 
    \includegraphics[width=0.8\columnwidth]{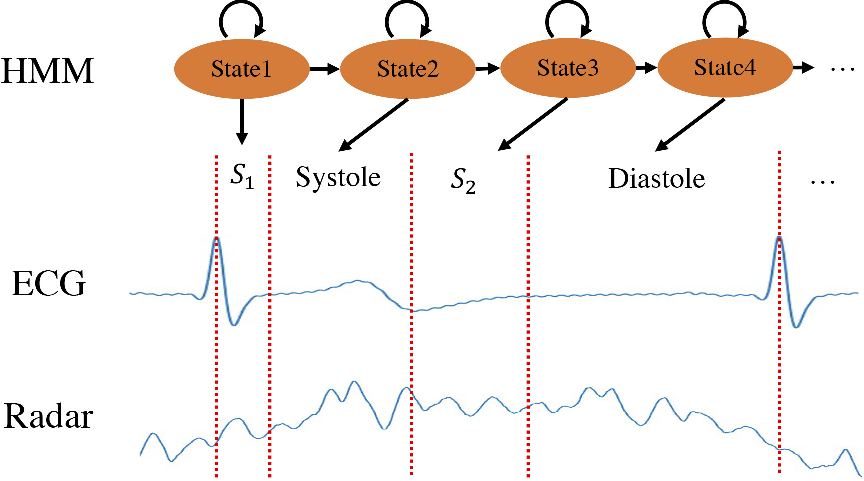}
    \caption{HMM for cardiac features segmentation on radar and ECG signals.} 
    \label{fig:hmm_princ} 
\end{figure}

Hidden Markov model (HMM) is a probabilistic model used to capture hidden states (e.g., the cardiac events) from observable sequential states and can be implemented for radar-based cardiac feature extraction with four states defined as $S_1$, systole, $S_2$ and diastole~\cite{shi2021contactless}, as illustrated in Figure~\ref{fig:hmm_princ}. Then, after training the HMM model using radar signals with labelled states, the hidden information (i.e., different states) can be first identified for the new inputs, enabling further cardiac feature extraction by calculating the time intervals between the identified states.

The prediction accuracy of HMM can be further improved with hidden semi-Markov model by taking the uneven duration of each state into consideration, making the probability of transition to a new state obey the nature of cardiac events~\cite{schmidt2010segmentation}. Furthermore, Xia \textit{et al.}~\cite{xia2021radar} first extracted the frequency envelogram of the radar signal between $8$ Hz and $50$ Hz using STFT and then applied the modified hidden semi-Markov model by introducing logistic regression to derive the emission and observation probability matrix, achieving discrimination between states. In \cite{shi2021contactless} and \cite{will2018radar}, hidden semi-Markov model was utilized to divide the radar-based heart sound signal with the aid of high-level features such as the Hilbert envelope and the power spectral density envelope to mitigate noises. Mei \textit{et al.}~\cite{mei2020fast} proposed a region-HMM that pre-set the effective frequency ranges of HR to estimate the HR with limited hidden states.

\subsection{Blind Source Separation Methods}
Blind source separation (BSS) is a classic problem originally in the audio processing field to separate a group of signals (e.g., voices of different people) from the mixed signal received by microphone(s)~\cite{cardoso1998blind}. For radar-based cardiac monitoring, the received signal is also a nonlinear mixing of various sources such as RR, RBM, cardiac vibration and multi-path interference~\cite{chen2021movi}. The methods introduced in this subsection aim to decompose the mixed signal and extract the cardiac vibration signal according to different criteria:
\begin{itemize}
    \item Multiple signal classification (MUSIC): orthogonal signal spaces for cardiac signal and noises;
    \item Independent component analysis (ICA): statistical independence between cardiac signal and noises;
    \item Empirical mode decomposition (EMD): time-scale features in mixed signals;
    \item Variational mode decomposition (VMD) and sparse signal reconstruction (SSR): natural sparsity of heartbeats.
\end{itemize}

\subsubsection{Multiple Signal Classification}
MUSIC is an eigenstructure-based method to decompose the signal into several orthogonal signal subspaces and assumes that the received signal is composed by a finite number ($P$) of sinusoidal signals~\cite{lee2016tracking}. MUSIC first calculates the covariance matrix for the mixed signal $f(n)$ and then applies eigenvalue decomposition (EVD) to decompose the covariance matrix into two orthogonal subspaces~\cite{lee2016tracking}. The $P$ largest eigenvalues are related to the signal subspace that contains cardiac features, while the rest of the eigenvectors form the noise subspace. To achieve the best performance, the value of $P$ should be selected carefully based on the number of sources in the ambient environment to achieve high frequency resolution using short window length~\cite{yamamoto2018non}; otherwise, the large $P$ may cause spurious peaks, and the small $P$ may bring the loss of some spectral components~\cite{bechet2013non}.

In literature, Lee \textit{et al.}~\cite{lee2016tracking} thought that although the dominant HR frequency could be ruined by motion artifacts or car vibration noise, the MUSIC algorithm with a fixed $P$ value could still reveal the HR harmonic to help the estimation of HR. Bechet \textit{et al.}~\cite{bechet2013non} re-selected the $P$ value before each measurement by evaluating the estimated error for $P\in [60,200]$, and the optimal $P$ value was selected as $140$ after the error evaluation. Before applying MUSIC, Yamamoto \textit{et al.}~\cite{yamamoto2018non} decomposed the signal by DCT to reveal the heartbeat-related components to help the estimation of $P$ values, decreasing time window to $5$ sec and achieving high estimation accuracy. Furthermore, to improve the sensitivity to the HRV, Yamamoto \textit{et al.}~\cite{yamamoto2019music} improved previous DCT-based MUSIC by repeatedly performing the MUSIC with a decreasing window length until there is only one peak left in a single time window.

\subsubsection{Independent Component Analysis}
\begin{figure}[tb] 
    \centering 
    \includegraphics[width=0.95\columnwidth]{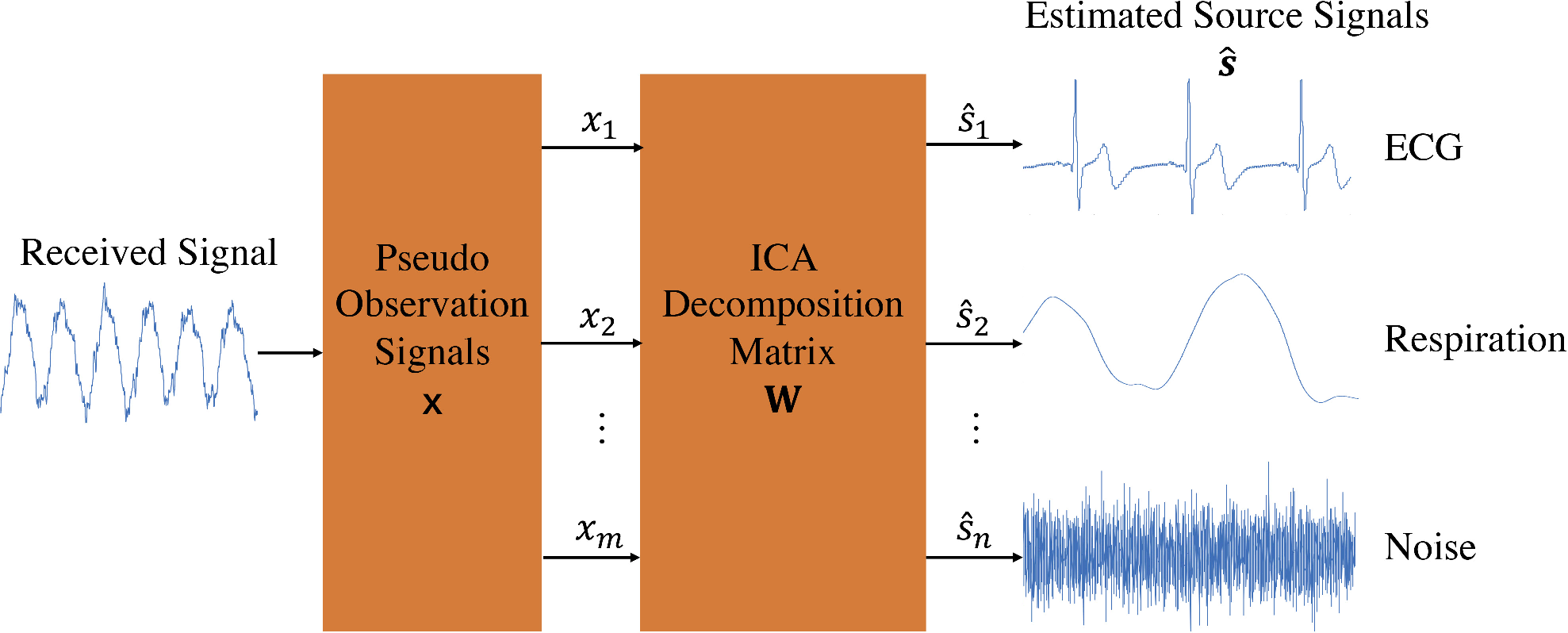}
    \caption{General procedure for cardiac feature extraction using ICA.} 
    \label{fig:ica} 
\end{figure}

ICA decomposes the mixed signal into several independent components with maximum statistical independence~\cite{ren2021vital} as shown in Figure~\ref{fig:ica} and can be mathematically expressed as:
\begin{equation}
    \mathbf{x}=\mathbf{A s}
\end{equation}
where $\mathbf{x}$ and $\mathbf{s}$ are $m$ observed signals and $n$ unknown source signals respectively and $\mathbf{A}$ is the matrix used to mix the source signals into the observed signals. The ICA algorithm
aims at estimating the decomposition matrix $\mathbf{W}$ to obtain the estimated source signals $\mathbf{\hat{s}}$ as:
\begin{equation}
\mathbf{\hat{s}}= \mathbf{W x}
\end{equation}
According to the central limit theorem~\cite{stone2004independent}, it is possible to obtain $\mathbf{W}$ by solving the optimization problem that maximizes the statistical independence of $\mathbf{\hat{s}}$. However, the ICA method requires the number of observed signals $m$ is not less than that of source signals $n$~\cite{mercuri2018direct}. Therefore, generating the uncorrelated (whiten) pseudo observation signals is compulsory for single-radar-based monitoring, as shown in Figure~\ref{fig:ica}. 

For the applications in literature, Mercur \textit{et al.}~\cite{mercuri2018direct} directly applied ICA using the pseudo signals with disjoint spectra generated by WT to enhance the chest movement in noisy scenarios. Furthermore, Mercur \textit{et al.} theoretically proved the capability of ICA to remove the multi-path interference in~\cite{mercuri2021enabling} and utilized singular value decomposition to remove the noise and generate the pseudo signals for the fast-ICA, simplifying the calculation of the statistical independence for source signals matrix. Lv \textit{et al.}~\cite{lv2021non} also used fast-ICA to suppress the mode aliasing effect, but with EMD method for pre-whitening and denoising. To reduce the noise, Xu \textit{et al.}~\cite{xu2021accurate} adopted fast-ICA to decompose the pre-whitened signals generated from EVD into several independent components, and the HR estimation was performed on specific independent components selected in terms of the signal frequency and power.

\subsubsection{Empirical Mode Decomposition}
\begin{figure}[tbp]
    \centering
    \subfloat[]{\label{fig:emd_examp}\includegraphics[width=0.5\columnwidth]{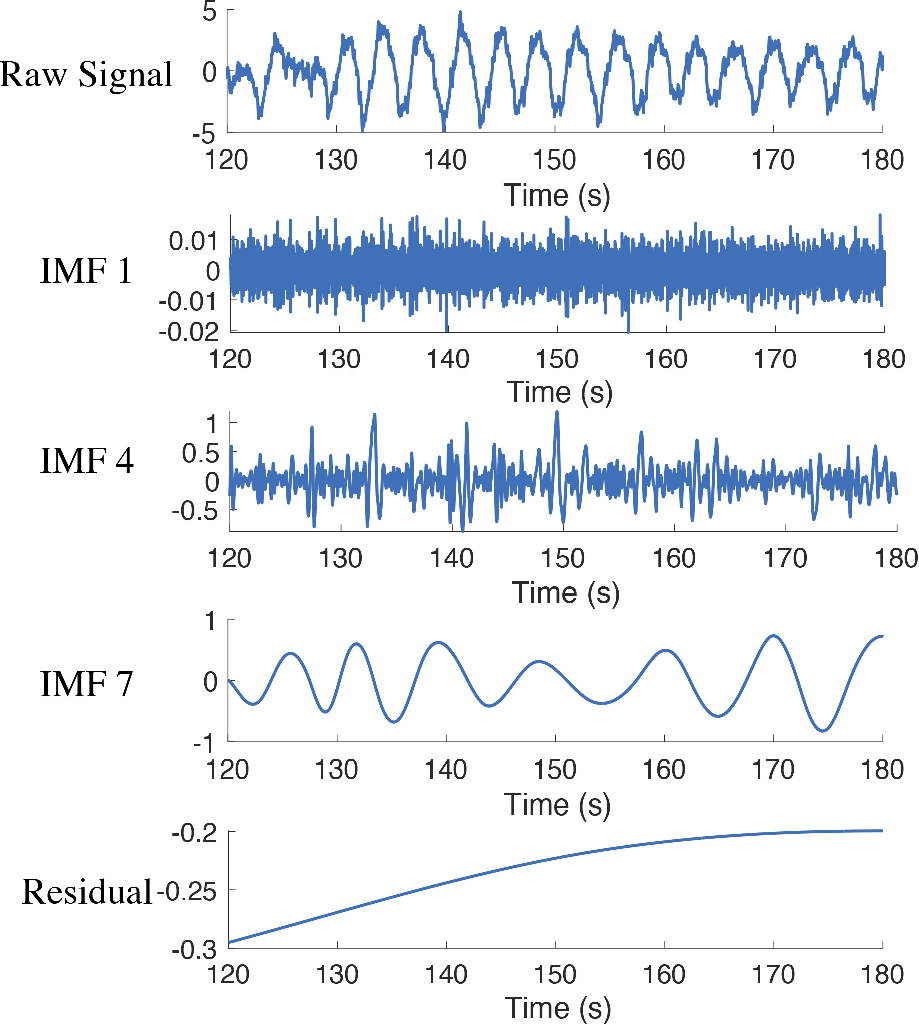}}
    \subfloat[]{\label{fig:emd_fft}\includegraphics[width=0.5\columnwidth]{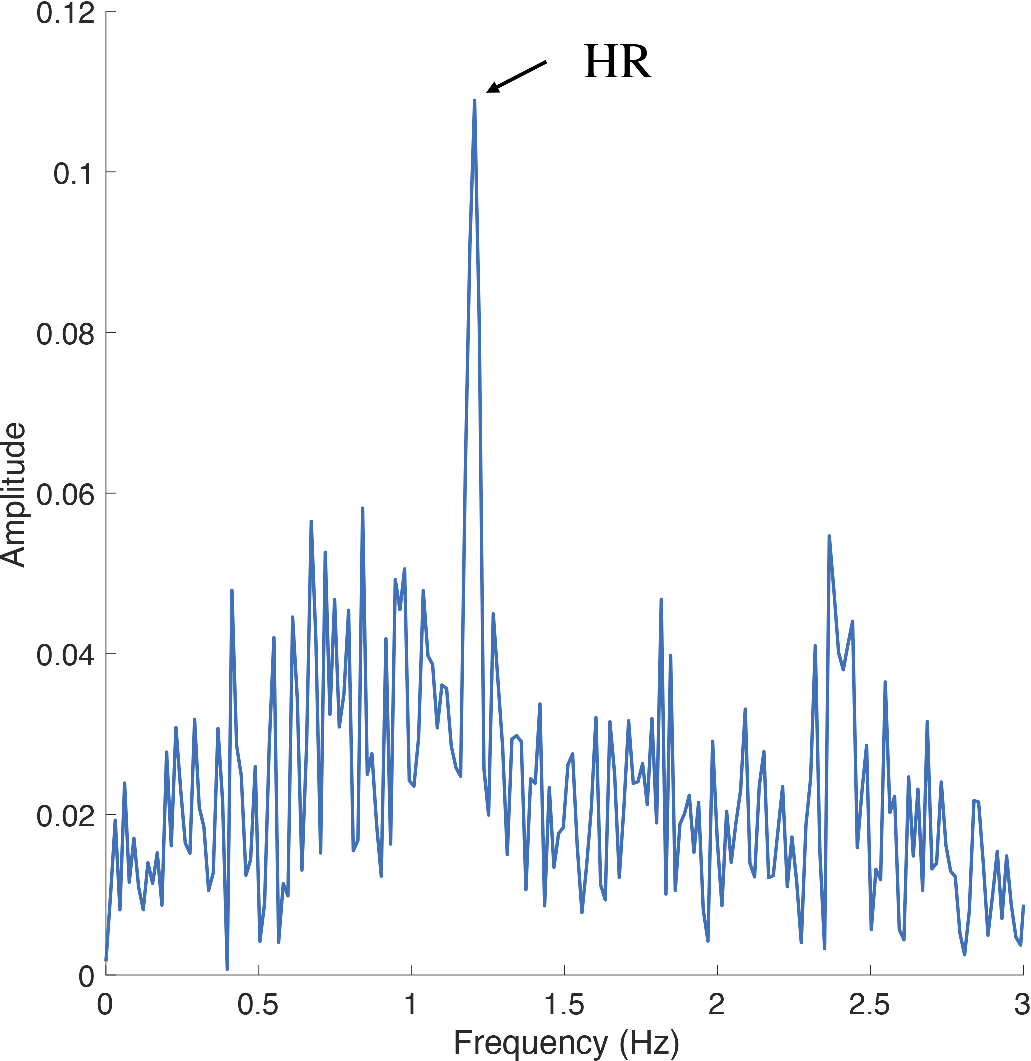}} 
    \caption{Illustration for EMD methods: (a) the raw radar signal is decomposed into nine IMFs (only three IMFs are shown here), with residual; (b) detected HR frequency from IMF $4$.}
    \label{fig:emd}
\end{figure}

Instead of using any wavelet basis, EMD directly extracts the local time-scale features by iteratively decomposing the mixed signal into multiple intrinsic mode functions (IMFs) with a residual. Therefore, EMD is capable of decomposing the non-linear modulated signals, whereas only some of these IMFs contain the cardiac information, as shown in Figure~\ref{fig:emd}. The conventional EMD method suffers from the mode mixing problem if the extremum of the signal is not uniformly distributed~\cite{zhang2020health} and may cause insufficient decomposition of the signal, especially for the low-frequency components. To avoid the mode mixing problem, Wu \textit{et al.}~\cite{weishaupt2018vital} proposed the ensemble EMD (EEMD) by adding the white Gaussian noise (WGN) with uniform frequency distribution in the mixed signal, because the mixture of the raw signal with WGN is continuous in time scale~\cite{zhang2020health}.

According to the literature, EEMD was used in~\cite{liu2020vital} to suppress the mutual radar interference with the IMF used for HR estimation selected according to the signal power and frequency. Another IMF selection strategy was proposed in~\cite{diraco2017radar} to reduce RBM noise by assigning a weight to each IMF to show the statistical difference between the current IMF and the raw mixed signal. Then, the best IMF can be selected by seeking the local minimum weight right after a local maximum because the weight would sharply decrease during the RBM.

For the further improvement of EEMD, Zhang \textit{et al.}~\cite{zhang2020health} claimed that the performance of EEMD highly depends on the choice of WGN~\cite{liang2018improved} because the WGN with high amplitude may distort the extreme distribution of low-frequency components and cause mode mixing, while the low-amplitude WGN may lead to the mode mixing for high-frequency components. Therefore, the adaptive EEMD was proposed with the amplitude of WGN changing as a sinusoidal function of the frequency to mitigate RBM~\cite{zhang2020health}. Furthermore, complete ensemble EMD with adaptive noise was used in~\cite{sun2020remote} to remove the residual noise brought by WGN and decrease the required iterations at the same time. To resist RBM and car vibrations, Shyu \textit{et al.}~\cite{shyu2018detection,shyu2020uwb} applied a two-layer EEMD with the first layer aiming to suppress the background scatter and align the fast-time-domain signals according to the energy function of the IMF. Then, the second layer was applied to the aligned signals in slow-time domain for the final HR estimation. 

\subsubsection{Variational Mode Decomposition}
Different from EMD with iterative calculations, VMD assumes the HR signal naturally has narrow bandwidth (sparsity on the spectrum). Therefore, it is possible to decompose the mixed signal into $K$ narrow-band IMFs in a non-recursive manner by ensuring: (a) the summation of the IMFs equals the raw mixed signal; (b) the total bandwidth of $K$ IMFs is minimum~\cite{dragomiretskiy2013variational}. Due to the restriction on the bandwidth, VMD can also overcome the mode mixing problem with the proper selection of $K$.

VMD was used in~\cite{shen2018respiration} with alternating direction method of multipliers as the optimizer to reveal the weak HR component from the chest movement signal. Similarly, Duan \textit{et al.}~\cite{duan2018non} used the same approach to decompose the signal, but the HR frequency is extracted from the Hilbert spectrum to reveal the change rule of the sequential data in real-time. In addition, similar to EMD, the performance of the VMD method highly relies on the choices of decomposition layer number $K$ and penalty value $\alpha$~\cite{zhang2021mutual}. Therefore, Zhang \textit{et al.}~\cite{zhang2021mutual} proposed to adaptively select the parameters according to the energy loss during the decomposition and then estimate the HR based on the mixed IMFs within the HR passband. Wang \textit{et al.}~\cite{wang2021mmhrv} applied a heuristic method to change $K$ and $\alpha$ iteratively, and the decomposition would be terminated once the HR components have been identified.

\subsubsection{Sparse Signal Reconstruction}
SSR method reconstructs the heartbeat signal from the mixed observed signals based on the natural sparsity of heartbeats in time-domain~\cite{zhang2014troika}. The spectrum of the reconstructed signal has a high resolution so that the frequency components of the heartbeat can be distinguished from those of the noises. The SSR model can be mathematically formulated as:
\begin{equation}
    \mathbf{y}=\mathbf{\Phi x}+\mathbf{v}
\end{equation}
where $y$ is the observed mixed signal, $\mathbf{x}$ is the solution signal with inherent sparsity (e.g., estimated HR signal), $\mathbf{v}$ is the noise, and $\mathbf{\Phi}$ represents a known basis matrix. For cardiac feature extraction, the basis matrix $\mathbf{\Phi}$ can be generally selected as Gaussian random matrix~\cite{candes2006stable}, and the goal of the SSR algorithm is to find the sparsest solution signal $\mathbf{x}$ based on $\mathbf{\Phi}$ and $\mathbf{y}$ by solving certain pre-defined optimization problem~\cite{luo2022pso,zeng2022pso}.

Based on the literature review, Ye \textit{et al.}~\cite{ye2018robust} first generated the sparse observed signal by singular spectrum analysis to remove the noises and then conducted SSR by the proposed zero-attracting sign-error least mean square algorithm. The zero-attracting algorithm was used to ensure the signal sparsity by combining the current error with the $\ell_{1}$-norm, and the sign-error least mean square algorithm was evolved from the least-mean-square adaptive filter by introducing the error with a sign to restrict the updating bound of error effectively. Furthermore, Ye \textit{et al.}~\cite{ye2018stochastic} improved zero-attracting sign-error least mean square algorithm by first applying an adaptive regularization parameter to achieve a proper weight of sparse penalty based on the standard deviation of the raw radar signal and then adopting the time-window-variation technique to enhance the HR component. Zhang \textit{et al.}~\cite{zhang2021mutual} utilized the improved morphological component analysis by finding the unique sparse basis and further suppressed the mutual radar interference based on the different amplitudes between target and inference signals. Similarly, Wang \textit{et al.}~\cite{wang2019noncontact} also tried to find the unique basis for the HR signal using the K-singular value decomposition algorithm by updating all the potential bases sequentially during the basis learning instead of updating them simultaneously. In addition, the idea of sparse penalty was also introduced to non-negative matrix factorization in~\cite{ye2019blind} by using sparse non-negative matrix factorization to replace singular spectrum analysis for separating the HR signal.

\subsection{Deep Learning Methods}
Deep learning belongs to machine learning based on artificial neural networks, but has multiple hidden layers formed by numerous neurons to achieve great performance. Each neuron receives a set of weighted inputs from other neurons and outputs the result through an activation function to introduce the non-linearity into the deep learning model~\cite{cheng2022deep}. Normally, the deep learning models with multiple hidden layers require complex structures and training methods, but are suitable for modelling the complex cardiac monitoring problem because the target signals are non-linearly modulated with other noises~\cite{chen2021movi}. In addition, to enable the deep learning model to learn the latent information (high-level features) buried in the signal, a massive amount of data mostly needs to be fed into the deep learning model to find the optimal weights after iterative training. Regarding the three main deep learning methods mentioned in the literature, CNN is famous for spatial feature identification, while LSTM is used for temporal feature extraction~\cite{nirmal2021deep}, and DCL can decompose the signal in a non-linear manner based on statistical independence~\cite{chen2021movi}. 

\subsubsection{Convolutional Neural Network}
\begin{figure}[tb] 
    \centering 
    \includegraphics[width=0.9\columnwidth]{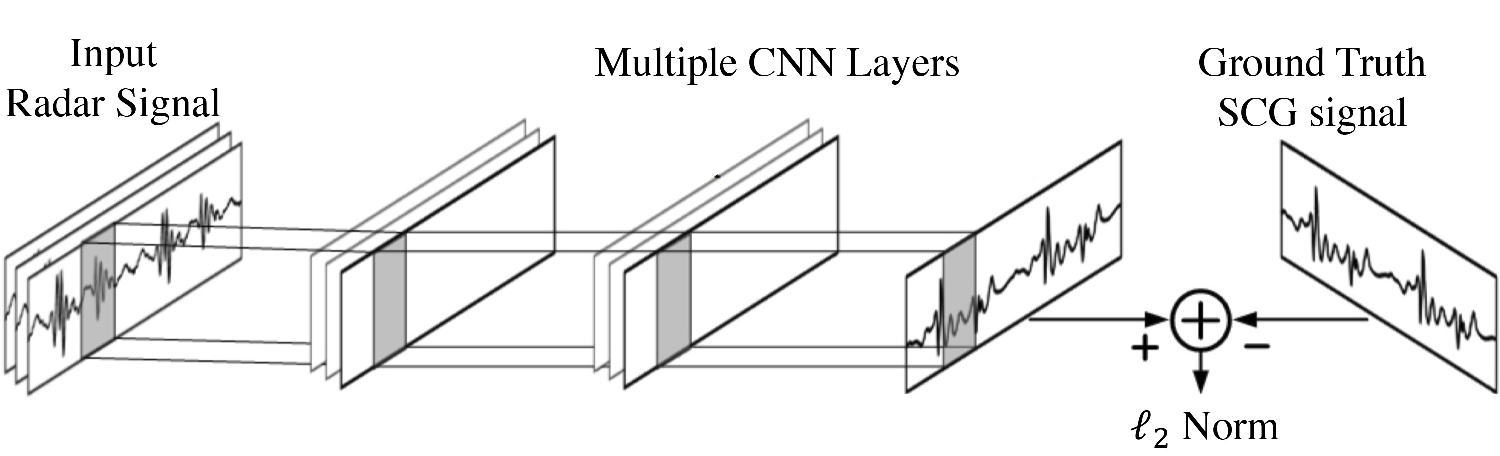}
    \caption{General CNN model for cardiac feature extraction with multiple CNN layers and ground truth cardiac feature (e.g., SCG signal)~\cite{ha2020contactless}} 
    \label{fig:cnn} 
\end{figure}

Due to the ability of non-linear projection~\cite{zhang2019end}, CNN-based algorithms are originally used for image processing to extract the high-level features by convolving the convolutional kernel with the input images, and recent research also applies CNN to the fine-grained cardiac monitoring (e.g., reconstruction of ECG or SCG signals). The common principle for CNN-based cardiac feature extraction is illustrated in Figure~\ref{fig:cnn} that several CNN layers are applied to extract the high-level feature embedded in input signals, and the expected output is the cardiac feature such as SCG or ECG signal. In addition, the model training is fulfilled by minimizing the loss function (e.g., $\ell_{2}$-norm) iteratively until convergence to maximize the similarity between the CNN output and the ground truth signal.

The aforementioned multiple-CNN-layer structure in Figure~\ref{fig:cnn} was adopted in~\cite{ha2020contactless} to imitate the effect of multiple FIR filters to extract different frequency components, and the output layer was used to combine these components to form the final SCG signal. Chen \textit{et al.}~\cite{chen2021movi} designed an encoder-decoder structure based on CNN to first extract the fine-grained features from the coarse HR signals and then cascade the signals obtained from consecutive windows together as the final output. Chen \textit{et al.}~\cite{chen2021contactless} added the novel transformer block after the CNN block to encode the temporal and spatial features, and the compressed features were then decoded by the temporal convolutional network to generate the fine-grained ECG output. Wu \textit{et al.}~\cite{wu2019person} thought that the ECG waveform contained too much redundant information for simple HR estimation. Therefore, the ground-truth signals were replaced by triangular signals with peaks representing the peaks in the ECG waveform. In addition, the CNN model was simultaneously trained by multiple independent CNN channels with time-delayed inputs to synchronize the clock between radar and ECG devices.

\subsubsection{Long Short-Term Memory Network}
Long short-term memory (LSTM) network evolves from the recurrent neural network~\cite{hochreiter1997long} by introducing the concept of ``gate'' to selectively remember or forget the information from previous time intervals. Due to the capability of memorizing the long-term dependence of the input signal, LSTM is widely applied in sequence-to-sequence translation~\cite{nirmal2021deep} and hence suitable for the radar-based cardiac feature extraction, which is essentially the translation from radar signal to cardiac feature signal.

Instead of using a sequential signal as input, Yamamoto \textit{et al.}~\cite{yamamoto2020ecg} adopted CNN-LSTM for ECG reconstruction by adding the CNN block before LSTM to accept the  signal spectrum as input, enabling the extraction of spatial and temporal features at the same time. Gong \textit{et al.}~\cite{gong2021rf} proposed self-calibrated LSTM to mitigate RBM and improve the generalization ability for the new participants under test. The proposed model leveraged the self-calibrated mechanism by taking the estimated HR for the static state as ground truth, and fine-tuning the parameters according to the estimation error for the RBM state of the new participants~\cite{IJSS-2021-06}. Another popular variant of LSTM is bidirectional-LSTM (Bi-LSTM), by adopting two LSTM layers and taking both forward and backward time dependency into account. Shi \textit{et al.}~\cite{shi2021contactless} applied Bi-LSTM to measure HRV induced by the cold pressor test, achieving a better performance than HMM method. Li \textit{et al.}~\cite{li2019standalone} utilized Bi-LSTM to classify the signal segments into positive or negative segments with a certain score, and the peaks from the score signal were finally used to identify the heartbeats. In addition, LSTM was used for signal decomposition in~\cite{ye2021spectral} by generating the embedding vectors for the input signals. After training by ground-truth ECG and RR signal spectra to learn the corresponding embedding vectors, the trained LSTM model could take radar signals as inputs and generate embedding vectors belonging to different clusters such as ECG, RR and noise. 

\subsubsection{Deep Contrastive Learning}
\begin{figure*}[tb] 
    \centering 
    \includegraphics[width=1.9\columnwidth]{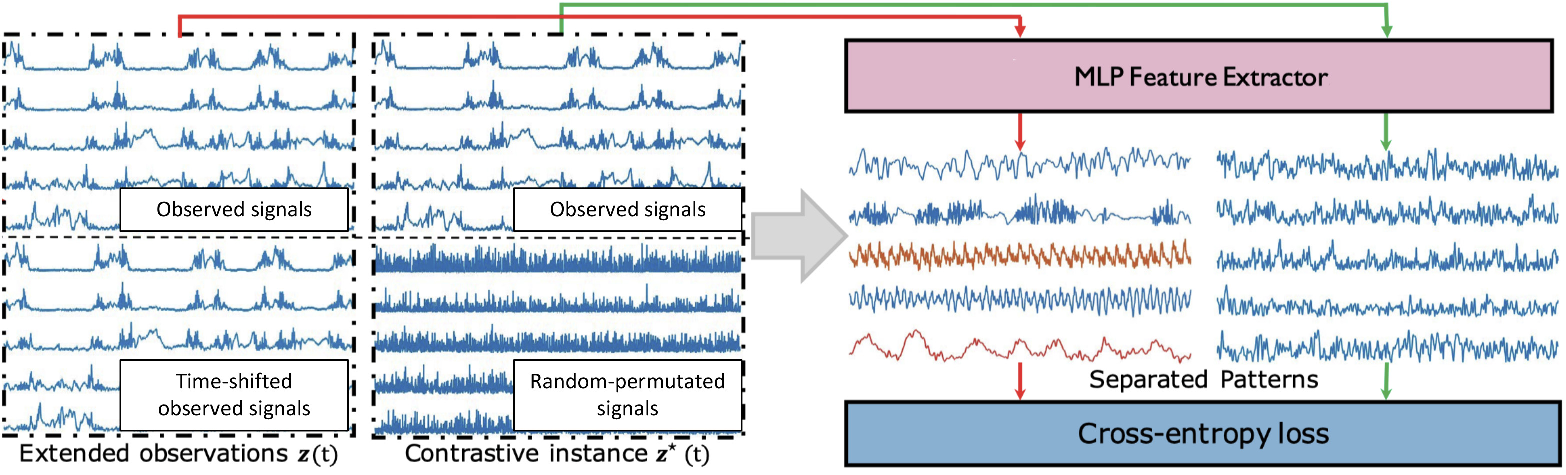}
    \caption{DCL method for cardiac feature extraction, with the left part illustrating the construction of positive samples (i.e., extended observations $z(\mathrm{t})$) and negative samples (i.e., contrastive instance $z^{\star}(\mathrm{t})$), and the right part showing the output of the trained multi-layer perceptron acting as a feature extractor~\cite{chen2021movi}.}
    \label{fig:dcl} 
\end{figure*}

Deep contrastive learning (DCL) is a self-supervised learning method that can learn the features without any ground-truth signal and is also able to map the non-linear transformation~\cite{chen2020simple}. The common principle of DCL is to first construct the positive and negative samples: the positive samples can be formed by the augmented version of the observed signals, and the negative samples can be formed by the random samples from the observed signal~\cite{jaiswal2020survey}. Then, the training stage of the DCL model aims to group the embeddings generated from the same sample while maximizing the distance between different samples.

To obtain the HR signal under the strong body movement, Chen \textit{et al.}~\cite{chen2021movi} applied the DCL method by firstly constructing the positive samples (i.e., extended observations $z(\mathrm{t})$) using the observed signals (i.e., original radar signal) and the time-shifted observed signals (i.e., original radar signal with a certain time lag) as shown in Figure~\ref{fig:dcl}. Similarly, the negative samples (i.e., contrastive instance $z^{\star}(\mathrm{t})$) are formed by the same observed radar signals as in $z(\mathrm{t})$ and the random-permutated signals as the noise signal. Therefore, the constructed $z^{\star}(\mathrm{t})$ has the same marginal distribution as $z(\mathrm{t})$ but with a heavily corrupted temporal structure, enabling the subsequent training of the feature extractor to extract cardiac features from the radar signal. In the second stage, the multi-layer perceptron model acting as a cardiac feature extractor was trained by minimizing the cross-entropy loss of a binary classification problem (positive/negative sample). After convergence, one of the output neurons can generate the coarse HR signal (shown as the orange signal in Figure~\ref{fig:dcl}), enabling the subsequent fine-grained HR estimation.

\begin{table*}[tb]
    \centering
    \caption{Datasets for cardiac monitoring using radar}
    \begin{tabular}{p{1.9cm}<{\centering} p{1.5cm} p{1.9cm}<{\centering} p{2.5cm} p{1.2cm}<{\centering} p{2.0cm} p{5cm} }
    \toprule
    \textbf{Reference} & \textbf{Radar Type} & \textbf{Frequency} & \textbf{Available Signals} & \textbf{Subjects} & \textbf{Data Length} & \textbf{Monitoring Configuration}\\
    \toprule
    Shi \textit{et al.}~\cite{shi2020dataset} 2020 & \makecell[tl]{Self-made \\ CW radar} & $24$ GHz & \makecell[tl]{1. Raw radar signal\\ \quad (I/Q channel)\\ 2. ECG and RR\\ 3. PCG } & $11$ Adults & \makecell[tl]{$20$ min/person\\ $3.7$ hours in total} & \makecell[tl]{1. Radar Position: Chest, Back, Carotid \\2. Radar Distance: $20$ - $30$ cm \\3. Body Position: Sitting \\4. Scenarios: Normal-breath, Holding-breath, \\\quad Post-exercise, Speaking } \\
    \toprule
    Schellenberger \textit{et al.}~\cite{schellenberger2020dataset} 2020 & \makecell[tl]{Self-made \\ CW radar} & $24$ GHz & \makecell[tl]{1. Raw radar signal\\ \quad (I/Q channel)\\2. ECG and RR\\3. PCG, ICG and BP$^{(1)}$} & $30$ Adults & \makecell[tl]{$50$ min/person\\$24$ hours in total} & \makecell[tl]{1. Radar Position: Chest \\2. Radar Distance: $40$ cm \\3. Body Position: Lying \\4. Scenarios: Normal-breath, Holding-breath, \\ \quad Valsalva manoeuvre, Tilt table test }  \\
    \toprule
    Yoo \textit{et al.}~\cite{yoo2021radar} 2021 & \makecell[tl]{TI$^{(2)}$ FMCW\\ IWR-$6843$ \\Tx1 Rx4$^{(3)}$} & \makecell[tl]{Start: $60$ GHz\\ BW$^{(4)}$: $4$ GHz} & \makecell[tl]{1. Raw radar signal\\2. HR and RR \\3. Heartbeat and \\ \quad breathing waveform} &$50$ Children & \makecell[tl]{$5$ min/person\\$4.1$ hours in total}  & \makecell[tl]{1. Radar Position: Chest \\2. Radar Distance:  N/A  \\3. Body Position: Sitting \\4. Scenarios: Normal-breath} \\
    \toprule
    Gong \textit{et al.}~\cite{gong2021rf} 2021 & \makecell[tl]{TI FMCW\\ AWR-$1843$} &\makecell[tl]{Start: $77$ GHz\\ BW: $4$ GHz} &\makecell[tl]{1. Raw radar signal \\2. HR and RR} & $14$ Adults & \makecell[tl]{$9$-$22$ min/person\\ Total hours: N/A} & \makecell[tl]{1. Radar Position: Whole body \\2. Radar Distance:  $1$ - $6$ m  \\3. Body Position: Standing \\4. Scenarios: Static, Walking, Exercise} \\
    \toprule
    Chen \textit{et al.}~\cite{chen2021contactless} 2021 & \makecell[tl]{TI FMCW\\ AWR-$1843$ \\Tx3 Rx4} &\makecell[tl]{Start: $77$ GHz\\ BW: $3.3$ GHz} &\makecell[tl]{1. Raw radar signal \\2. ECG} & $35$ Adults & \makecell[tl]{$17$ min/person\\$10$ hours in total} & \makecell[tl]{1. Radar Position: Chest \\2. Radar Distance:  $40$ - $50$ cm  \\3. Body Position: Lying \\4. Scenarios: Normal-breath, Irregular-breath,\\ \quad Post-exercise, Sleeping } \\
    \toprule
    Edanami \textit{et al.}~\cite{edanami2022medical} 2022 & \makecell[tl]{NJR$^{(5)}$ CW \\ NJR-$4262$ \\ NJR-$4178$J}& $24$ and $10$ GHz & \makecell[tl]{1. Raw radar signal \\ \quad (I/Q for $24$ GHz, \\\quad I for $10$ GHz) \\2. ECG and RR } & $9$ Adults & \makecell[tl]{$10$ min/person\\$1.5$ hours in total} & \makecell[tl]{1. Radar Position: Back (under bed) \\2. Radar Distance:  $15$ cm  \\3. Body Position: Lying \\4. Scenarios: Normal-breath } \\
    \bottomrule
    \multicolumn{7}{l}{(1) PCG: Phonocardiography; ICG: Impedance Cardiogram; BP: Blood Pressure; (2) TI: Texas Instruments;} \\ 
    \multicolumn{7}{l}{(3) Tx1 Rx4: $1$ Transmitter and $4$ Receivers; (4) BW: Bandwidth; (5) NJR: New Japan Radio.}
    \end{tabular}
    \label{tab:dataset}
\end{table*}

\section{Review of Public Datasets}\label{sec:dataset}
For radar-based cardiac monitoring, researchers tend to collect their own small-scale datasets because most methods mentioned in the last section do not require pre-training on a dataset. In contrast, the recently emerged deep learning-based methods normally rely on massive data to train and evaluate the proposed deep learning model. However, the trend of using radar with high operating frequency raises high demand on the radar architecture design (e.g., high-speed analog-to-digital converter)~\cite{li2013review} and noise cancellation algorithms (e.g., to mitigate multi-path effect)~\cite{mercuri2021enabling}, increasing the complexity of data collection. Therefore, the release of the public datasets can not only help the new researchers start their work or verify/compare the proposed algorithms in different scenarios, but also could benefit the companies to design the template or framework for cardiac feature extraction in developing the commercial radar platforms. This section first lists the public datasets containing synchronous radar signals and ground-truth cardiac feature signals (with key information of data collection provided in Table~\ref{tab:dataset}), and then evaluates each dataset from the perspective of potential usage on different scenarios or algorithms.

\begin{itemize}
    \item \textbf{Shi \textit{et al.}~\cite{shi2020dataset}:} This dataset contains not only the ground-truth ECG measurements but also the heart sound (i.e. phonocardiography (PCG)) signals, enabling the research of radar-based heart sound detection. In addition, the radar is faced towards three different body regions with different incident angles during recording, helping researchers to investigate cardiac monitoring with arbitrary radar positions.
    \item \textbf{Schellenberger \textit{et al.}~\cite{schellenberger2020dataset}: }This huge dataset has $24$ hours of raw radar signal and abundant ground-truth signals. In addition, several special testing procedures are designed (such as Valsalvathe manoeuvre and tilt table test) to induce a more abrupt HRV compared with the variation caused by intensive exercise. Therefore, the sensitivity of the cardiac monitoring algorithms to detect HRV can be evaluated by this dataset.
    \item \textbf{Yoo \textit{et al.}~\cite{yoo2021radar}: }This dataset is collected from children under $13$ years old and originally used for radar-based age-group classification. The dataset only provides coarse HR information, and the data is all recorded for children with normal breath in a noiseless scenario. Therefore, this dataset is friendly to new researchers.
    \item \textbf{Gong \textit{et al.}~\cite{gong2021rf}:} This dataset is collected by FMCW radar with high operating frequency under various scenarios. Especially, the walking scenario is considered during the data collection, and the radar is oriented to the whole human body instead of a specific area with a short monitoring distance, enabling future research about monitoring from a long distance or under vigorous body movement.
    \item \textbf{Chen \textit{et al.}~\cite{chen2021contactless}:} This dataset adopts FMCW radar with high operating frequency and long data length for deep learning-based research. In addition, the data collection is performed under various scenarios to induce HRV or introduce extra noise, enabling the potential development of the generally applicable cardiac monitoring algorithms in complex scenarios.
    \item \textbf{Edanami \textit{et al.}~\cite{edanami2022medical}: }This is a small dataset with only $1.5$ hours of data collected from people with normal breath by two radars positioned under the bed with different operating frequencies. Therefore, the research based on this dataset can verify the feasibility of the future smart bed with vital sign monitoring function~\cite{zhai2022contactless}, or investigate the multi-radar-based cardiac monitoring techniques.
\end{itemize}
        
In summary, some of the public datasets (such as Yoo \textit{et al.}~\cite{yoo2021radar} and Edanami \textit{et al.}~\cite{edanami2022medical}) contain short data records collected in a single scenario under mild noises, and are suitable for new researchers to test the algorithms performance (e.g., real-time ability of spectrum-based methods) or validate potentially effective algorithms in an early design stage. Additionally, some public datasets (such as Chen \textit{et al.}~\cite{chen2021contactless} and Schellenberger \textit{et al.}~\cite{schellenberger2020dataset}) contain long data records with large HRV or chest wall displacement as might be encountered in real-world cardiac monitoring, enabling the development of complex algorithms (e.g. parameters selection for BSS methods or model training for deep learning methods) in realistic scenarios. The release of these public datasets benefits the new researchers because collecting the private dataset is labour-intensive work and requires extensive radar and cardiac monitoring knowledge. In addition, the research based on public datasets can avoid the duplication of similar datasets and benefit the comparison between different algorithms. However, the number of existing public datasets is still insufficient to support various research interests, such as mitigating severe RBM and monitoring cardiac features in the narrow space. Therefore, future researchers are encouraged to produce high-quality public datasets under ample scenarios to benefit the development of cardiac monitoring techniques. 
        
\section{Challenges and Future Work}\label{sec:candf}
Although the algorithms introduced in Section~\ref{sec:algos} show the feasibility of radar-based cardiac feature extraction, there still exist many obstacles that prevent the application of cardiac monitoring in real-world scenarios. This section will first evaluate the pros and cons of the algorithms in different categories as also summarized in Table~\ref{tab:procon}, and then several unsolved challenges will be discussed with possible solutions as future directions.

\subsection{Algorithms Evaluation}

\begin{table*}[tb]
    \centering
    \caption{Advantages and disadvantages of cardiac monitoring algorithms} 
    \begin{tabular}{l p{6.1cm} p{6.1cm}}
    \toprule
    \textbf{Category} & \textbf{Advantage} & \textbf{Disadvantage}\\
    \toprule
    \textbf{Spectrum-based Methods} &\makecell[tl]{1. Have low algorithm complexity\\ 2. Can be implemented directly on the hardware \\ 3. Good real-time performance}   & \makecell[tl]{1. Hardly detect the abrupt HRV \\2. Easily affected by noise} \\
    \toprule
    \textbf{Periodicity-based Methods} &\makecell[tl]{1. Can identify individual heartbeat \\ 2. Can realize cardiac event segmentation\\3. Naturally immune to non-periodic noise} &\makecell[tl]{1. Require prior knowledge \\ 2. Hardly detect rarely happened cardiac events\\ 3. Require calibration for new participants} \\
    \toprule
    \textbf{Blind Source Separation Methods} &\makecell[tl]{1. No need for prior knowledge \\ 2. Can perform non-linear or non-stationary signal \\\quad decomposition}  & \makecell[tl]{1. Require a careful selection of the parameters and \\\quad the resultant decomposed signals \\ 2. May insufficiently or over decompose the mixed \\\quad signals} \\
    \toprule
    \textbf{Deep Learning Methods} & \makecell[tl]{1. Can reconstruct the fine-grained cardiac features\\ \quad (e.g., ECG waveform) \\ 2. Can leverage long-term dependency in estimation \\ 3. Can potentially resist different kinds of noises} & \makecell[tl]{1. Require massive data for training\\ 2.  Rely on full-featured dataset with numerous\\ \quad scenarios for good performance \\ 3. Use large deep learning model and cannot be \\\quad implemented on lightweight devices}\\
    \bottomrule
    \end{tabular}%
    \label{tab:procon}%
\end{table*}%

\subsubsection{Spectrum-Based Methods}
The spectrum-based methods are matured in the signal-processing area and can be implemented directly on the hardware with limited computational resources while achieving good real-time performance~\cite{hu2014real,park2017polyphase}. In fact, several commercial data-capture boards for raw radar signal processing have already embedded these algorithms on FPGA~\cite{obadi2021survey}. 

For the disadvantages, the spectrum-based methods are not sensitive to the abrupt HRV, because multiple cardiac cycles are involved in a single segment truncated by the time window, but only one HR value can be estimated from each segment~\cite{shi2021contactless}. Another problem caused by the truncation is that each segment may not contain integral multiples of the cardiac cycles, causing the shift of the estimated HR around the true HR~\cite{ye2018stochastic}. Lastly, various noises (such as RR harmonics and slight RBM~\cite{yamamoto2018spectrogram}) may have the frequency components falling into the HR frequency band and hence dominate the HR components~\cite{rong2019remote}. Although it is possible to extract the cardiac features using HR harmonics from the high-frequency spectrum (e.g., $1.5-5$ Hz) to avoid the effect of strong RR harmonic components~\cite{le2019multivariate}, the HR harmonics sometimes might be too weak and easily distorted by unexpected noises (e.g., car vibrations~\cite{da2019theoretical}).

\subsubsection{Periodicity-Based Methods}
The periodicity-based methods take the natural periodicity of cardiac features into account and could identify a single heartbeat by either measuring the similarity between template and raw signals or predicting the next heartbeat based on a probabilistic model. In addition, periodicity-based methods further enable the fine cardiac event segmentation which is valuable for clinical diagnosis~\cite{ha2020contactless}. Lastly, periodicity-based methods are naturally immune to the noises that do not show any periodic feature.

The above mentioned advantages of periodicity-based methods rely on certain prior knowledge (e.g., BBI or ECG waveform) to design the template/model, but such template/model may be unsuitable for the monitoring of new participants and hence require certain calibration during actual monitoring~\cite{gong2021rf}. Furthermore, the optimal template/model with periodicity features learned from datasets may not fit the diverse cardiac features of other individuals~\cite{zhang2020health}. In other words, the researchers need to balance the ability to detect rare cardiac events (e.g., heart diseases) against the ability to provide accurate HR estimations for most scenarios.

\subsubsection{Blind Source Separation Methods}
BSS methods do not require prior knowledge regarding cardiac events and can decompose the mixed signal according to specific characteristics. In addition, instead of modelling the heartbeat signal as a summation of a finite number of periodic sinusoidal signals~\cite{shen2018respiration}, most BSS methods assume that the source signals are not simple sinusoidal functions but with narrow bandwidth, enabling the non-linear decomposition of the mixed signal.

For the drawbacks, an ideal decomposition result relies on the careful selection of the parameters (e.g., $P$ for MUSIC, WGN for EMD, $K$ for VMD), whereas these parameters are obtained empirically and require a re-selection after altering the monitoring environment. The inappropriate selection of parameters can cause the mode mixing or over-decomposition issue, making it impossible to extract HR from the decomposed signals~\cite{weishaupt2018vital}. Lastly, it is still a challenge to select the proper decomposed signals for HR estimation because there are multiple decomposed signals falling into the heartbeat frequency band~\cite{shyu2018detection}.

\subsubsection{Deep Learning Methods}
Due to the powerful fitting capability of the deep learning model, deep learning methods can model complex and non-linear projections and hence produce fine-grained cardiac signals compared with the algorithms in other categories. In addition, with a special network design (e.g., LSTM network), the deep learning model can memorize long-term dependency from the dataset, making stable long-term cardiac monitoring possible~\cite{IJSS-2021-03}. Lastly, compared with the other three categories of methods, the deep learning methods can potentially resist different kinds of noises by introducing these noises during the dataset collection~\cite{chen2021movi}.

The outstanding performance of deep learning methods generally relies on the training with a large-scale and full-featured dataset~\cite{schellenberger2020dataset}, whereas the well-trained deep learning model may fail to infer on the data out of distribution~\cite{gong2021rf}. For example, the deep learning model trained on the dataset for normal people cannot perform accurate cardiac monitoring for patients with different cardiac features. Similarly, the deep learning model trained on the dataset collected for multi-path effect mitigation cannot resist the RBM noise. Therefore, the collected dataset should contain various real-world scenarios, increasing the complexity of establishing a full-featured dataset. Additionally, it is still a challenge to implement complex deep learning models on compact and lightweight devices with real-time monitoring ability, because effective deep learning models normally require ample computational resources and large memory.

\subsubsection{Robustness Evaluation}
In addition to the pros/cons evaluations in terms of novelty or accuracy of certain algorithms, it is also necessary to especially analyse the robustness to show the feasibility and commercialization prospect of certain algorithms in real-world scenarios:

\begin{itemize}
    \item Some spectrum-based methods, such as FFT, BPF, auto-correlation and peak detection~\cite{zhang2021mutual}, have already been embedded on the highly encapsulated commercial radar platform~\cite{lee2018novel,lee2020feasibility}, or the radar-on-chip platform (e.g., Texas Instruments AWR-$1843$~\cite{chen2021contactless}). Although achieving good real-time performance on the small-scale device with limited computational resources, most spectrum-based methods are not robust to real-world noise (e.g., slight RBM) due to the lack of algorithm complexity and can only perform coarse cardiac monitoring in constrained scenarios (e.g., static or breath-holding people~\cite{park2021preclinical,lee2018novel}). Fortunately, the commercial radar-on-chip platforms enable researchers to test techniques (e.g., time-window variation technique~\cite{tu2015fast} or high-frequency spectrum evaluation~\cite{yamamoto2018spectrogram}) for future commercialization, and hence improve the robustness of spectrum-based methods while maintaining a good real-time ability using limited computational resources~\cite{hu2014real,park2017polyphase}.
    \item Periodicity-based methods leverage prior knowledge about cardiac features and are naturally robust to the noises without any periodic feature (e.g., large-scale RBM~\cite{lv2018doppler}), whereas sacrificing the robustness with respect to periodic noises~\cite{xu2021accurate} and the detection of rare cardiac events~\cite{zhang2020health}.
    \item BSS methods not only can mitigate the RR noise and slight RBM due to the nonlinear decomposition ability, but also show the potential to deal with complex noises such as mutual-radar interference and multi-path propagation~\cite{mercuri2021enabling,zhang2021mutual,liu2020vital}. However, more experiments are required to prove the robustness of BSS methods in real-world scenarios instead of only in the laboratory.
    \item Deep learning methods can potentially be robust in many hard cases, such as large-scale RBM~\cite{chen2021movi}, long-distance monitoring~\cite{gong2021rf}, fine-grained cardiac feature monitoring~\cite{ha2020contactless} and irregular cardiac event detection~\cite{shi2021contactless}. However, it is hard to simultaneously contain the aforementioned cases into a single dataset and further deploy the deep learning model on the commercial radar platform with limited computational resources.
\end{itemize}

In summary, many spectrum-based methods have been proven to have good real-time ability and low computational resource requirements~\cite{hu2014real,park2017polyphase}, and are promising to improve the robustness of the current commercial radar platform immediately. In addition, to realize robust cardiac monitoring in complex scenarios, new platforms with functional modules (e.g., neural network processing unit) need to be designed to support algorithms from the other three categories. Lastly, the feasibility of a new algorithm for future commercial radar platforms depends on the trade-off among algorithm complexity, required computational resources and desired monitoring accuracy/robustness.

\subsection{Challenges}\label{sec:chall}

\subsubsection{Body Movement}
Although the effect of body movement has been an unsolved challenge for a long time since the start of research in this area, most of the algorithms introduced above can only tolerate slight or abrupt body movement. In other words, consistent or strong body movements can still significantly distort the radar signal and affect the cardiac feature extraction~\cite{gong2021rf}. In addition, the cardiac monitoring for moving subjects is also challenging because the proposed system should be able to track the individual person spatially~\cite{mercuri2019vital}.

Some researchers tried to leverage other sensors or extra radio frequency tags to get the body movement information and compensate the phase signals accordingly~\cite{cardillo2021vital}. Otherwise, without the assistance from extra sensors, only deep learning-based methods are potentially capable of measuring the HR under vigorous movements such as walking and abrupt standing up~\cite{gong2021rf,chen2021movi}. However, the current work can only mitigate vigorous RBM under certain pre-defined modes (e.g., walking on the treadmill), therefore, is not generally applicable to mitigate the RBM under various unpredictable modes as in real-world scenarios.

\subsubsection{Complex Electromagnetic Environment}
The current cardiac monitoring experiments are normally performed under a complex electromagnetic environment full of electronic devices, and the radio frequency signals emitted by these devices can interfere with the radar used for cardiac monitoring. Additionally, the radar with a high operating frequency may need to address the multi-path effect, especially for the future implementation in limited space (e.g., in-cabin applications)~\cite{zhang2021mutual}.

Although mutual-radar interference and multi-path propagation have been investigated a lot in the radar area, relevant research in the cardiac monitoring area is still rare because most of the experiments are performed in a simple environment with the prominent reflection from human body. According to the literature, the signal models for mutual-radar interference and multi-path propagation were deduced in~\cite{mercuri2021enabling} and \cite{zhang2021mutual} respectively, whereas the corresponding experiments were performed under mild electromagnetic noises (e.g., interference generated by single radar or multi-path propagation in a large room). Therefore, it is necessary to conduct the experiment in a real complex electromagnetic environment (such as in-cabin or multi-radar scenarios) to verify the feasibility of cardiac monitoring.

\subsubsection{Long-Range Monitoring}
Long-range monitoring is critical for future applications (e.g., smart home) but will attenuate the reflected power from the body while enhancing the background clutter~\cite{cardillo2021vital}, causing poor cardiac monitoring accuracy. Therefore, most researchers position the radar within a range of $0.3-2$ meters to guarantee a high-SNR received signal. Some researchers try to enlarge the monitoring distance by designing advanced radar architectures, whereas the effective designs normally require massive antenna or complex transceiver module~\cite{feng2021multitarget,lin2022broadband}, and hence are incompatible with many real-world applications that require compact devices or real-time monitoring ability. Therefore, it is worth investigating the algorithms, such as digital beamforming~\cite{chen2021contactless}, for enhancing vital sign signals from a long-range measurement with compact radar platforms.

Another problem caused by long distance is the interruption during monitoring. For example, the hand movement in front of the chest can disturb monitoring~\cite{ha2020contactless}, or other people may break into the sight of the radar during the test. The current compromise is to first detect such interruptions based on the abrupt increase of the signal power and then replace the corrupted estimation with previous measurements~\cite{ha2020contactless}, whereas the monitoring will be terminated if the interruption exists for too long. Therefore, advanced algorithms should be developed to perform cardiac monitoring in the presence of either temporary or permanent interruptions.
        
\subsubsection{Radar Orientation and Self-Movement}
It has been verified in many papers that the optimal radar orientation is towards the chest area for the best SNR, and the second optimum is towards the back~\cite{ren2021vital}. However, the ability of cardiac monitoring from any orientation is crucial for some future applications such as smart home. Distributed radar network can be a promising solution because the orientation of a single radar cannot be guaranteed in real-world applications\cite{ren2021vital, IJSS-2021-05}.

The radar system installed on an unstable platform can also affect the SNR, because the extra platform vibration noise caused by radar self-movement (RSM) cannot be reckoned as white Gaussian noise and is hard to be removed~\cite{da2019theoretical,IJNDI-2022-09}. Although some researchers claimed that their methods are applicable to the unstable radar platform~\cite{shyu2020uwb,jia2018wifind}, few of them considered these noises in experiments. One possible solution requires introducing extra sensors (e.g., accelerometer)~\cite{IJSS-2021-08} to provide vibration or RSM information for phase compensation~\cite{da2019theoretical}. Otherwise, it is still a challenge to resist RSM with a single radar for cardiac feature monitoring. For inspiration, Cardillo \textit{et al.}~\cite{cardillo2021vital} identified the RSM according to the signals reflected from background clutters using a single radar, but only the respiration signal was reconstructed, and the proposed method cannot deal with large RSM. 

\subsubsection{Heart Rate Variation and Heart Rate Range}\label{sec:hrv}
The ability of HRV detection is crucial to cardiac monitoring for medical diagnosis, because patients with heart diseases (e.g., arrhythmia) normally have a large/abrupt HRV~\cite{gouveia2022bio}, destroying the intrinsic patterns of cardiac events. In the literature, few studies aimed to detect large/abrupt HRV because most monitoring algorithms rely on the intrinsic features of cardiac events. For instance, Shi \textit{et al.}~\cite{shi2021contactless} tried to induce the HRV through experiments such as cold pressor test and reported that the time-series model (e.g., HMM and LSTM) could potentially detect the HRV. However, the induced HRV is still different from the real data collected from patients in terms of cardiac morphology~\cite{sai2022cognitive}. The current research on HRV detection is mainly restricted by the limited data collected from real patients, and future studies may apply data augmentation or transfer learning to overcome this challenge.

For the heart rate range, almost all the researchers pre-determine a passband for the interested cardiac features. Current studies preferred a narrow passband (e.g., $1-1.6$ Hz) to include few in-band noise components. However, human HR can easily exceed $1.6$ Hz after mild exercise, while extending the objective passband raises demand for the noise robustness of the algorithms. Therefore, it is necessary to develop cardiac monitoring techniques that can measure a wide HR range while resisting possible noises.

\subsection{Future Works}\label{sec:future}
\subsubsection{Deep Learning for Spatiotemporal Cardiac Feature Extraction}
The cardiac events show the unique characteristics in both frequency and time domains as mentioned before, while most deep learning-based research only focuses on extracting either spatial or temporal features using a single deep learning model from radar signal. To extract the spatiotemporal features at the same time, it is possible to combine various deep learning blocks in series or parallel as used in image or speech processing~\cite{zhang2019end}. In addition, some advanced deep learning techniques, such as the attention mechanism, can be directly added to conventional deep learning models to help understand the global context~\cite{guan2022man}, improving the ability of cardiac feature extraction and noise robustness. Lastly, it is worth pre-processing the raw radar signal with other methods (e.g., denoising by EEMD) to reduce the burden of the deep learning model. In the future, the researchers could either design complex frameworks for cardiac feature extraction and signal decomposition based on functional deep learning blocks, or directly apply some open-source deep learning frameworks to process non-linearly mixed radar signals~\cite{SSCE-2022-02}.

\subsubsection{Data Augmentation for Abnormal Cardiac Features}
Numerous real-world applications (e.g., medical monitoring and driver monitoring systems) require the ability to detect abnormal cardiac features (e.g., arrhythmia), whereas most algorithms are unable to reveal such features, as discussed in Section~\ref{sec:hrv}. Although deep learning methods show the potential for HRV detection, the lack of abnormal cardiac data restricts the further development of advanced deep learning algorithms~\cite{shi2021contactless}. Therefore, it is necessary to augment the training data especially for abnormal cardiac features, while the current research only adopts simple augmentation techniques such as overlapping, stretching, noising and applying random polynomials~\cite{ha2021wistress}. In future studies, the advanced data augmentation methods can be helpful to even generate valuable data for abnormal cardiac events based on the real-collected datasets~\cite{SSCE-2021-01}.
        
\subsubsection{Transfer Learning from Other Cardiac Signals}
Transfer learning is a promising method to solve the inaccurate cardiac feature extraction caused by individual differences and insufficient data~\cite{chen2022msdan,wan2021review}. Despite the lack of radar-ECG data, numerous datasets containing pure ECG or PCG-ECG signals are available, with the data for patients also included. Transfer learning can extract the features from a similar domain and provide the prior knowledge to assist the training in another domain~\cite{wan2022eeg}, but the current research only transfers the model pre-trained by pure ECG dataset to fulfil the classification task for diagnosis purpose~\cite{yin2019self}. In the future, the deep learning model is expected to be first pre-trained by the dataset collected for similar tasks (e.g., extracting ECG from PCG signals). Then, with further training by a small amount of radar-ECG data, the transfer learning can adjust the pre-trained model to fit the radar-based cardiac feature extraction task~\cite{IJNDI-2022-04}.

\subsubsection{Evaluation Metrics for Fine-Grained Cardiac Feature Extraction}
Many algorithms require to evaluate the fidelity of the reconstructed signal for fine-grained cardiac monitoring. In literature, the mean squared error is mostly used to measure the global average distance between the ground truth and the estimated signal. However, the mean squared error tends to flatten the peaks and valleys of the signal~\cite{xu2021cardiacwave}, which are the key features for medical diagnosis. Therefore, new evaluation metrics need to be proposed to preserve the key features of cardiac signals, enabling further evaluation of the reconstructed signals from the medical perspective.

\subsubsection{SNR Improvement before Cardiac Feature Extraction}
Instead of developing various advanced signal processing algorithms, researchers are recommended to improve the SNR of the received signal before applying any cardiac feature extraction algorithm to reduce the burden of the subsequent processing algorithms, because the coarse heartbeat can be directly measured from the high-SNR signal without extra denoising or signal decomposition~\cite{ha2020contactless}. Additionally, the introduction of SNR-improvement techniques does not need to consider the nature of cardiac activities, and researchers can leverage the advanced developments from other areas to benefit radar-based cardiac monitoring, especially for the long-range monitoring scenario. One approach to increase SNR is to investigate the antenna design such as phased array radar or massive multiple-input multiple-output radar~\cite{feng2021multitarget}. Another approach is to leverage the FMCW radar platform and design the digital beamforming techniques to only extract the heartbeat-related signals from certain range bins~\cite{ha2020contactless,chen2021contactless}. In addition, programmable wireless environment is a recently emerged approach to improve SNR by leveraging artificial metamaterial, and can help mitigate the multi-path propagation or manipulate the stray Wi-Fi signal for vital sign monitoring as suggested in~\cite{li2019intelligent}.

\section{Conclusions}\label{sec:conclusions}
Radar-based contactless cardiac monitoring is promising to replace the contact-based monitoring method in future applications such as smart home and in-cabin monitoring. This review mainly focuses on the algorithms for cardiac feature extraction based on the signals collected from radar and proposes a new taxonomy to reveal the core features of these algorithms, helping the researchers and practitioners understand the principle and suitable application of each algorithm. In addition, five public datasets are listed and evaluated to help the new researchers start their work. Lastly, the pros and cons of the reviewed algorithms are discussed with the unsolved challenges and possible future directions. Furthermore, this review provides some insights into the development of cardiac monitoring: (a) the early studies mainly focused on radar architecture design and baseband signal processing, while recently many researchers are steering toward investigating the cardiac feature extraction algorithms to extract fine-grained cardiac features, mitigate real-world noises, realize multi-person monitoring or enable multi-sensor/radar-based monitoring; (b) deep learning methods receive much attention due to the outstanding performance, while the further improvement may require the large-scale and full-featured datasets which is hard to be collected. In the future, the researchers are encouraged to solve the challenges proposed in this review and produce high-quality public datasets, helping radar-based contactless cardiac monitoring truly benefit human society.


\begin{thebibliography}{100}

\bibitem{krittanawong2021integration}
C.~Krittanawong, A.~J. Rogers, K.~W. Johnson, Z.~Wang, M.~P. Turakhia, J.~L. Halperin, and S.~M. Narayan, ``Integration of novel monitoring devices with machine learning technology for scalable cardiovascular management,'' \emph{Nature Reviews Cardiology}, vol.~18, no.~2, pp. 75--91, Feb. 2021.

\bibitem{chen2018arrhythmia}
W.~Zhu, X.~Chen, Y.~Wang, L.~Wang, ``Arrhythmia recognition and classification using ECG morphology and segment feature analysis,'' \emph{IEEE/ACM Transactions on Computational Biology and Bioinformatics}, vol.~16, no.~1, pp. 131--138, Jun. 2018.

\bibitem{chen2021contactless}
J.~Chen, D.~Zhang, Z.~Wu, F.~Zhou, Q.~Sun, and Y.~Chen, ``Contactless electrocardiogram monitoring with millimeter wave radar,'' \emph{IEEE Transactions on Mobile Computing}, pp.~1--17, Oct. 2022.

\bibitem{ye2021spectral}
C.~Ye and T.~Ohtsuki, ``Spectral viterbi algorithm for contactless wide-range heart rate estimation with deep clustering,'' \emph{IEEE Transactions on Microwave Theory and Techniques}, vol.~69, no.~5, pp. 2629--2641, May 2021.

\bibitem{singh2020multi}
A.~Singh, S.~U. Rehman, S.~Yongchareon, and P.~H.~J. Chong, ``{Multi-resident non-contact vital sign monitoring using radar: A review},'' \emph{IEEE Sensors Journal}, vol.~21, no.~4, pp. 4061--4084, Feb. 2021.

\bibitem{jia2018wifind}
W.~Jia, H.~Peng, N.~Ruan, Z.~Tang, and W.~Zhao, ``{WiFind: Driver fatigue detection with fine-grained Wi-Fi signal features},'' \emph{IEEE Transactions on Big Data}, vol.~6, no.~2, pp. 269--282, Jun. 2020.

\bibitem{adib2015smart}
F.~Adib, H.~Mao, Z.~Kabelac, D.~Katabi, and R.~C. Miller, ``Smart homes that monitor breathing and heart rate,'' in \emph{Proceedings of the 33rd Annual ACM Conference on Human Factors in Computing Systems}, pp. 837--846, Apr. 2015.

\bibitem{kang2013various}
H.-B. Kang, ``Various approaches for driver and driving behavior monitoring: A review,'' in \emph{Proceedings of the IEEE International Conference on Computer Vision Workshops (ICCVW)}, pp. 616--623, Dec. 2013.

\bibitem{shen2018respiration}
H.~Shen, C.~Xu, Y.~Yang, L.~Sun, Z.~Cai, L.~Bai, E.~Clancy, and X.~Huang, ``{Respiration and heartbeat rates measurement based on autocorrelation using IR-UWB radar},'' \emph{IEEE Transactions on Circuits and Systems II: Express Briefs}, vol.~65, no.~10, pp. 1470--1474, Oct. 2018.

\bibitem{chen2018video}
X.~Chen, J.~Cheng, R.~Song, Y.~Liu, R.~Ward, and Z.~J. Wang, ``{Video-based heart rate measurement: Recent advances and future prospects},'' \emph{IEEE Transactions on Instrumentation and Measurement}, vol.~68, no.~10, pp. 3600--3615, Nov. 2018.  

\bibitem{chian2022vital}
D.-M. Chian, C.-K. Wen, C.-J. Wang, M.-H. Hsu, and F.-K. Wang, ``{Vital signs identification system with Doppler radars and thermal camera},'' \emph{IEEE Transactions on Biomedical Circuits and Systems}, vol.~16, no.~1, pp. 153--167, Feb. 2022.

\bibitem{xu2020leveraging}
X.~Xu, J.~Yu, and Y.~Chen, ``Leveraging acoustic signals for fine-grained breathing monitoring in driving environments,'' \emph{IEEE Transactions on Mobile Computing}, vol.~21, no.~3, pp. 1018--1033, Mar. 2020.

\bibitem{feng2021multitarget}
C.~Feng, X.~Jiang, M.-G. Jeong, H.~Hong, C.-H. Fu, X.~Yang, E.~Wang, X.~Zhu, and X.~Liu, ``{Multitarget vital signs measurement with chest motion imaging based on MIMO radar},'' \emph{IEEE Transactions on Microwave Theory and Techniques}, vol.~69, no.~11, pp. 4735--4747, Nov. 2021.

\bibitem{nirmal2021deep}
I.~Nirmal, A.~Khamis, M.~Hassan, W.~Hu, and X.~Zhu, ``Deep learning for radio-based human sensing: Recent advances and future directions,'' \emph{IEEE Communications Surveys \& Tutorials}, vol.~23, no.~2, pp. 995--1019, 2nd Quart. 2021.

\bibitem{xia2021radar}
W.~Xia, Y.~Li, and S.~Dong, ``Radar-based high-accuracy cardiac activity sensing,'' \emph{IEEE Transactions on Instrumentation and Measurement}, vol.~70, pp. 1--13, Jan. 2021.

\bibitem{wang2020remote}
Y.~Wang, W.~Wang, M.~Zhou, A.~Ren, and Z.~Tian, ``{Remote monitoring of human vital signs based on 77-GHz mm-wave FMCW radar},'' \emph{Sensors}, vol.~20, no.~10, p. 2999, May 2020.

\bibitem{islam2022contactless}
S.~M. Islam, O.~Boric-Lubecke, V.~M. Lubecke, A.-K. Moadi, and A.~E. Fathy, ``Contactless radar-based sensors: Recent advances in vital-signs monitoring of multiple subjects,'' \emph{IEEE Microwave Magazine}, vol.~23, no.~7, pp. 47--60, Jul. 2022.

\bibitem{wang2010novel}
F.-K. Wang, C.-J. Li, C.-H. Hsiao, T.-S. Horng, J.~Lin, K.-C. Peng, J.-K. Jau, J.-Y. Li, and C.-C. Chen, ``A novel vital-sign sensor based on a self-injection-locked oscillator,'' \emph{IEEE Transactions on Microwave Theory and Techniques}, vol.~58, no.~12, pp. 4112--4120, Dec. 2010.

\bibitem{droitcour2004range}
A.~D. Droitcour, O.~Boric-Lubecke, V.~M. Lubecke, J.~Lin, and G.~T. Kovacs, ``{Range correlation and I/Q performance benefits in single-chip silicon Doppler radars for noncontact cardiopulmonary monitoring},'' \emph{IEEE Transactions on Microwave Theory and Techniques}, vol.~52, no.~3, pp. 838--848, Mar. 2004.

\bibitem{ha2020contactless}
U.~Ha, S.~Assana, and F.~Adib, ``Contactless seismocardiography via deep learning radars,'' in \emph{Proceedings of the 26th Annual International Conference on Mobile Computing and Networking (MobiCom)}, pp. 1--14, Apr. 2020.

\bibitem{ji2022rbhhm}
S.~Ji, Z.~Zhang, Z.~Xia, H.~Wen, J.~Zhu, and K.~Zhao, ``RBHHM: A novel remote cardiac cycle detection model based on heartbeat harmonics,'' \emph{Biomedical Signal Processing and Control}, vol.~78, p. 103936, Sep. 2022.

\bibitem{ha2021wistress}
U.~Ha, S.~Madani, and F.~Adib, ``Wistress: Contactless stress monitoring using wireless signals,'' in \emph{Proceedings of the ACM on Interactive, Mobile, Wearable and Ubiquitous Technologies (IMWUT)}, vol.~5, no.~3, pp. 1--37, Sep. 2021.

\bibitem{yamamoto2020ecg}
K.~Yamamoto, R.~Hiromatsu, and T.~Ohtsuki, ``{ECG signal reconstruction via Doppler sensor by hybrid deep learning model with CNN and LSTM},'' \emph{IEEE Access}, vol.~8, pp. 130\,551--130\,560, Jul. 2020.

\bibitem{mercuri2021enabling}
M.~Mercuri, Y.~Lu, S.~Polito, F.~Wieringa, Y.-H. Liu, A.-J. van~der Veen, C.~Van~Hoof, and T.~Torfs, ``Enabling robust radar-based localization and vital signs monitoring in multipath propagation environments,'' \emph{IEEE Transactions on Biomedical Engineering}, vol.~68, no.~11, pp. 3228--3240, Nov. 2021.

\bibitem{zhang2021mutual}
X.~Zhang, Z.~Liu, Y.~Kong, and C.~Li, ``Mutual interference suppression using signal separation and adaptive mode decomposition in noncontact vital sign measurements,'' \emph{IEEE Transactions on Instrumentation and Measurement}, Dec. 2021.

\bibitem{chen2021movi}
Z.~Chen, T.~Zheng, C.~Cai, and J.~Luo, ``{MoVi-Fi: Motion-robust vital signs waveform recovery via deep interpreted RF sensing},'' in \emph{Proceedings of the 27th Annual International Conference on Mobile Computing and Networking (MobiCom)}, pp. 392--405, Feb. 2021.

\bibitem{lee2016tracking}
K.~J. Lee, C.~Park, and B.~Lee, ``{Tracking driver's heart rate by continuous-wave Doppler radar},'' in \emph{2016 38th Annual International Conference of the IEEE Engineering in Medicine and Biology Society (EMBC)}, pp. 5417--5420, Aug. 2016.

\bibitem{gong2021rf}
J.~Gong, X.~Zhang, K.~Lin, J.~Ren, Y.~Zhang, and W.~Qiu, ``{RF Vital Sign Sensing Under Free Body Movement},'' \emph{Proceedings of the ACM on Interactive, Mobile, Wearable and Ubiquitous Technologies (IMWUT)}, vol.~5, no.~3, pp. 1--22, Sep. 2021.

\bibitem{tomii2015heartbeat}
S.~Tomii and T.~Ohtsuki, ``{Heartbeat detection by using Doppler radar with wavelet transform based on scale factor learning},'' in \emph{2015 IEEE International Conference on Communications (ICC)}, pp. 483--488, Jun. 2015.

\bibitem{zhang2020health}
J.~Zhang, Y.~Wu, Y.~Chen, and T.~Chen, ``Health-radio: Towards contactless myocardial infarction detection using radio signals,'' \emph{IEEE Transactions on Mobile Computing}, vol.~21, no.~2, pp. 585--597, Feb. 2022.

\bibitem{mogi2017heartbeat}
E.~Mogi and T.~Ohtsuki, ``{Heartbeat detection with Doppler radar based on spectrogram},'' in \emph{2017 IEEE International Conference on Communications (ICC)}, pp. 1--6, May 2017.

\bibitem{zhao2016emotion}
M.~Zhao, F.~Adib, and D.~Katabi, ``Emotion recognition using wireless signals,'' in \emph{Proceedings of the 22nd Annual International Conference on Mobile Computing and Networking (MobiCom)}, pp. 95--108, Oct. 2016.

\bibitem{lv2018doppler}
Q.~Lv, L.~Chen, K.~An, J.~Wang, H.~Li, D.~Ye, J.~Huangfu, C.~Li, and L.~Ran, ``Doppler vital signs detection in the presence of large-scale random body movements,'' \emph{IEEE Transactions on Microwave Theory and Techniques}, vol.~66, no.~9, pp. 4261--4270, Sep. 2018.

\bibitem{diraco2017radar}
G.~Diraco, A.~Leone, and P.~Siciliano, ``A radar-based smart sensor for unobtrusive elderly monitoring in ambient assisted living applications,'' \emph{Biosensors}, vol.~7, no.~4, p.~55, Nov. 2017.

\bibitem{shyu2018detection}
K.~Shyu, L.~Chiu, P.~Lee, T.~Tung, and S.~Yang, ``{Detection of breathing and heart rates in UWB radar sensor data using FVPIEF-based two-layer EEMD},'' \emph{IEEE Sensors Journal}, vol.~19, no.~2, pp. 774--784, Jan. 2019.

\bibitem{shyu2020uwb}
K.~Shyu, L.~Chiu, P.~Lee, and L.~Lee, ``{UWB simultaneous breathing and heart rate detections in driving scenario using multi-feature alignment two-layer EEMD method},'' \emph{IEEE Sensors Journal}, vol.~20, no.~17, pp. 10\,251--10\,266, Sep. 2020.

\bibitem{ye2018stochastic}
C.~Ye, K.~Toyoda, and T.~Ohtsuki, ``{A stochastic gradient approach for robust heartbeat detection with Doppler radar using time-window-variation technique},'' \emph{IEEE Transactions on Biomedical Engineering}, vol.~66, no.~6, pp. 1730--1741, Jun. 2018.

\bibitem{ye2018robust}
C.~Ye, K.~Toyoda, and T.~Ohtsuki, ``{Robust heartbeat detection with Doppler radar based on stochastic gradient approach},'' in \emph{2018 IEEE International Conference on Communications (ICC)}, pp. 1--6, May 2018.

\bibitem{ye2019blind}
C.~Ye, K.~Toyoda, and T.~Ohtsuki, ``Blind source separation on non-contact heartbeat detection by non-negative matrix factorization algorithms,'' \emph{IEEE Transactions on Biomedical Engineering}, vol.~67, no.~2, pp. 482--494, Feb. 2020.

\bibitem{xiong2017accurate}
Y.~Xiong, S.~Chen, X.~Dong, Z.~Peng, and W.~Zhang, ``{Accurate measurement in Doppler radar vital sign detection based on parameterized demodulation},'' \emph{IEEE Transactions on Microwave Theory and Techniques}, vol.~65, no.~11, pp. 4483--4492, Mar. 2017.

\bibitem{xiong2020differential}
Y.~Xiong, Z.~Peng, C.~Gu, S.~Li, D.~Wang, and W.~Zhang, ``Differential enhancement method for robust and accurate heart rate monitoring via microwave vital sign sensing,'' \emph{IEEE Transactions on Instrumentation and Measurement}, vol.~69, no.~9, pp. 7108--7118, Sep. 2020.

\bibitem{tu2015fast}
J.~Tu and J.~Lin, ``Fast acquisition of heart rate in noncontact vital sign radar measurement using time-window-variation technique,'' \emph{IEEE Transactions on Instrumentation and Measurement}, vol.~65, no.~1, pp. 112--122, Jan. 2016.

\bibitem{kwon2021attention}
H.~B. Kwon, S.~H. Choi, D.~Lee, D.~Son, H.~Yoon, M.~H. Lee, Y.~J. Lee, and K.~S. Park, ``{Attention-based LSTM for non-Contact sleep stage classification using IR-UWB radar},'' \emph{IEEE Journal of Biomedical and Health Informatics}, vol.~25, no.~10, pp. 3844--3853, Oct. 2021.

\bibitem{nosrati2019concurrent}
M.~Nosrati, S.~Shahsavari, S.~Lee, H.~Wang, and N.~Tavassolian, ``{A concurrent dual-beam phased-array Doppler radar using MIMO beamforming techniques for short-range vital-signs monitoring},'' \emph{IEEE Transactions on Antennas and Propagation}, vol.~67, no.~4, pp. 2390--2404, Jan. 2019.

\bibitem{mercuri2019vital}
M.~Mercuri, I.~R. Lorato, Y.-H. Liu, F.~Wieringa, C.~V. Hoof, and T.~Torfs, ``Vital-sign monitoring and spatial tracking of multiple people using a contactless radar-based sensor,'' \emph{Nature Electronics}, vol.~2, no.~6, pp. 252--262, Jun. 2019.

\bibitem{wang2020multiple}
F.-K. Wang, P.-H. Juan, D.-M. Chian, and C.-K. Wen, ``Multiple range and vital sign detection based on single-conversion self-injection-locked hybrid mode radar with a novel frequency estimation algorithm,'' \emph{IEEE Transactions on Microwave Theory and Techniques}, vol.~68, no.~5, pp. 1908--1920, May 2020.

\bibitem{adib20143d}
F.~Adib, Z.~Kabelac, D.~Katabi, and R.~C. Miller, ``{3D tracking via body radio reflections},'' in \emph{11th USENIX Symposium on Networked Systems Design and Implementation (NSDI)}, pp. 317--329, Apr. 2014.

\bibitem{ren2021vital}
W.~Ren, F.~Qi, F.~Foroughian, T.~Kvelashvili, Q.~Liu, O.~Kilic, T.~Long, and A.~E. Fathy, ``Vital sign detection in any orientation using a distributed radar network via modified independent component analysis,'' \emph{IEEE Transactions on Microwave Theory and Techniques}, vol.~69, no.~11, pp. 4774--4790, Nov. 2021.

\bibitem{gravina2017multi}
R.~Gravina, P.~Alinia, H.~Ghasemzadeh, and G.~Fortino, ``Multi-sensor fusion in body sensor networks: State-of-the-art and research challenges,'' \emph{Information Fusion}, vol.~35, pp. 68--80, May 2017.

\bibitem{8281483}
I.~D. Castro, M.~Mercuri, T.~Torfs, I.~Lorato, R.~Puers, and C.~Van~Hoof, ``{Sensor Fusion of Capacitively Coupled ECG and Continuous-Wave Doppler Radar for Improved Unobtrusive Heart Rate Measurements},'' \emph{IEEE Journal on Emerging and Selected Topics in Circuits and Systems}, vol.~8, no.~2, pp. 316--328, Feb. 2018.

\bibitem{bruser2015ambient}
C.~Br{\"u}ser, C.~H. Antink, T.~Wartzek, M.~Walter, and S.~Leonhardt, ``Ambient and unobtrusive cardiorespiratory monitoring techniques,'' \emph{IEEE Reviews in Biomedical Engineering}, vol.~8, pp. 30--43, Mar. 2015.

\bibitem{wang2022multisensor}
H.~Wang and Y.~Liu, ``Multisensor data fusion for life detection in cluttered environments,'' \emph{IEEE Sensors Journal}, vol.~22, no.~24, pp. 24\,559--24\,566, Nov. 2022.

\bibitem{bai2018fusion} 
X.~Bai, Z.~Wang, L.~Sheng, and Z.~Wang, ``{Reliable data fusion of hierarchical wireless sensor networks with asynchronous measurement for greenhouse monitoring},'' \emph{IEEE Transactions on Control Systems Technology}, vol.~27, no.~3, pp.~1036--1046, Feb. 2018.

\bibitem{gupta2022automatic}
K.~Gupta, M.~Srinivas, J.~Soumya, O.~J. Pandey, and L.~R. Cenkeramaddi, ``{Automatic contact-less monitoring of breathing rate and heart rate utilizing the fusion of mmWave radar and camera steering system},'' \emph{IEEE Sensors Journal}, vol.~22, no.~22, pp. 22\,179--22\,191, Oct. 2022.

\bibitem{gu2013hybrid}
C.~Gu, G.~Wang, Y.~Li, T.~Inoue, and C.~Li, ``A hybrid radar-camera sensing system with phase compensation for random body movement cancellation in Doppler vital sign detection,'' \emph{IEEE transactions on microwave theory and techniques}, vol.~61, no.~12, pp. 4678--4688, Nov. 2013.

\bibitem{li2013review}
C.~Li, V.~M. Lubecke, O.~Boric-Lubecke, and J.~Lin, ``{A review on recent advances in Doppler radar sensors for noncontact healthcare monitoring},'' \emph{IEEE Transactions on Microwave Theory and Techniques}, vol.~61, no.~5, pp. 2046--2060, May 2013.

\bibitem{obadi2021survey}
A.~B. Obadi, P.~J. Soh, O.~Aldayel, M.~H. Al-Doori, M.~Mercuri, and D.~Schreurs, ``{A survey on vital signs detection using radar techniques and processing with FPGA implementation},'' \emph{IEEE Circuits and Systems Magazine}, vol.~21, no.~1, pp. 41--74, Feb. 2021.

\bibitem{lin1975noninvasive}
J.~C. Lin, ``Noninvasive microwave measurement of respiration,'' \emph{Proceedings of the IEEE}, vol.~63, no.~10, pp. 1530--1530, Oct. 1975.

\bibitem{wang2020experimental}
D.~Wang, S.~Yoo, and S.~H. Cho, ``Experimental comparison of IR-UWB radar and FMCW radar for vital signs,'' \emph{Sensors}, vol.~20, no.~22, p. 6695, Nov. 2020.

\bibitem{hirt2003ultra}
W.~Hirt, ``Ultra-wideband radio technology: Overview and future research,'' \emph{Computer Communications}, vol.~26, no.~1, pp. 46--52, Jan. 2003.

\bibitem{huang2021antennas}
Y.~Huang, \emph{Antennas: From theory to practice}.\hskip 1em plus 0.5em minus 0.4em\relax New York, NY, USA: John Wiley \& Sons, 2021.

\bibitem{ramasubramanian2018moving}
K.~Ramasubramanian and K.~Ramaiah, ``{Moving from legacy 24 GHz to state-of-the-art 77-GHz radar},'' \emph{ATZelektronik Worldwide}, vol.~13, no.~3, pp. 46--49, Jun. 2018.

\bibitem{obeid2008low}
D.~Obeid, G.~Issa, S.~Sadek, G.~Zaharia, and G.~El~Zein, ``{Low power microwave systems for heartbeat rate detection at 2.4, 5.8, 10 and 16 GHz},'' in \emph{2008 First International Symposium on Applied Sciences on Biomedical and Communication Technologies}, pp. 1--5, Oct. 2008.

\bibitem{zhou2022towards}
Y.~Zhou, L.~Liu, H.~Zhao, M.~L{\'o}pez-Ben{\'\i}tez, L.~Yu, and Y.~Yue, ``Towards deep radar perception for autonomous driving: Datasets, methods, and challenges,'' \emph{Sensors}, vol.~22, no.~11, p. 4208, May 2022.

\bibitem{petrovic2019high}
V.~L. Petrovi{\'c}, M.~M. Jankovi{\'c}, A.~V. Lup{\v{s}}i{\'c}, V.~R. Mihajlovi{\'c}, and J.~S. Popovi{\'c}-Bo{\v{z}}ovi{\'c}, ``{High-accuracy real-time monitoring of heart rate variability using 24 GHz continuous-wave Doppler radar},'' \emph{IEEE Access}, vol.~7, pp. 74\,721--74\,733, Jun. 2019.

\bibitem{saluja2019supervised}
J.~Saluja, J.~Casanova, and J.~Lin, ``{A supervised machine learning algorithm for heart-rate detection using Doppler motion-sensing radar},'' \emph{IEEE Journal of Electromagnetics, RF and Microwaves in Medicine and Biology}, vol.~4, no.~1, pp. 45--51, Mar. 2020.

\bibitem{yang2020vital}
Z.-K. Yang, H.~Shi, S.~Zhao, and X.-D. Huang, ``{Vital sign detection during large-scale and fast body movements based on an adaptive noise cancellation algorithm using a single Doppler radar sensor},'' \emph{Sensors}, vol.~20, no.~15, p. 4183, Jul. 2020.

\bibitem{zhu2018fundamental}
F.~Zhu, K.~Wang, and K.~Wu, ``{A fundamental-and-harmonic dual-frequency Doppler radar system for vital signs detection enabling radar movement self-cancellation},'' \emph{IEEE Transactions on Microwave Theory and Techniques}, vol.~66, no.~11, pp. 5106--5118, Nov. 2018.

\bibitem{le2020heartbeat}
M.~Le, ``Heartbeat extraction based on a high order derivative for ultra-wideband impulse radar application,'' \emph{Journal of Physics D: Applied Physics}, vol.~53, no.~18, p. 18LT02, Feb. 2020.

\bibitem{hu2014real}
W.~Hu, H.~Zhang, Z.~Zhao, Y.~Wang, and X.~Wang, ``{Real-time remote vital sign detection using a portable Doppler sensor system},'' in \emph{2014 IEEE Sensors Applications Symposium (SAS)}, pp. 89--93, Feb. 2014.

\bibitem{park2017polyphase}
J.~Park, J.-W. Ham, S.~Park, D.-H. Kim, S.-J. Park, H.~Kang, and S.-O. Park, ``{Polyphase-basis discrete cosine transform for real-time measurement of heart rate with CW Doppler radar},'' \emph{IEEE Transactions on Microwave Theory and Techniques}, vol.~66, no.~3, pp. 1644--1659, Mar. 2018.

\bibitem{shih2021quadrature}
J.-Y. Shih and F.-K. Wang, ``{Quadrature cosine transform (QCT) with varying window length (VWL) technique for noncontact vital sign monitoring using a continuous-wave (CW) radar},'' \emph{IEEE Transactions on Microwave Theory and Techniques}, Mar. 2022.

\bibitem{li2017wavelet}
M.~Li and J.~Lin, ``{Wavelet-transform-based data-length-variation technique for fast heart rate detection using 5.8-GHz CW Doppler radar},'' \emph{IEEE Transactions on Microwave Theory and Techniques}, vol.~66, no.~1, pp. 568--576, Jan. 2017.

\bibitem{liu2022vital}
S.~Liu, Q.~Qi, H.~Cheng, L.~Sun, Y.~Zhao, and J.~Chai, ``{A Vital Signs Fast Detection and Extraction Method of UWB Impulse Radar Based on SVD},'' \emph{Sensors}, vol.~22, no.~3, p. 1177, Feb. 2022.

\bibitem{ling2022non}
Z.~Ling, W.~Zhou, Y.~Ren, J.~Wang, and L.~Guo, ``Non-contact heart rate monitoring based on millimeter wave radar,'' \emph{IEEE Access}, vol.~10, pp. 74\,033--74\,044, Jul. 2022.

\bibitem{kim2019peak}
J.-Y. Kim, J.-H. Park, S.-Y. Jang, and J.-R. Yang, ``{Peak detection algorithm for vital sign detection using Doppler radar sensors},'' \emph{Sensors}, vol.~19, no.~7, p. 1575, Apr. 2019.

\bibitem{yamamoto2018spectrogram}
K.~Yamamoto, K.~Toyoda, and T.~Ohtsuki, ``{Spectrogram-based non-contact RRI estimation by accurate peak detection algorithm},'' \emph{IEEE Access}, vol.~6, pp. 60\,369--60\,379, Oct. 2018.

\bibitem{xu2021accurate}
H.~Xu, M.~P. Ebrahim, K.~Hasan, F.~Heydari, P.~Howley, and M.~R. Yuce, ``Accurate heart rate and respiration rate detection based on a higher-order harmonics peak selection method using radar non-contact sensors,'' \emph{Sensors}, vol.~22, no.~1, p.~83, Dec. 2021.

\bibitem{will2018radar}
C.~Will, K.~Shi, S.~Schellenberger, T.~Steigleder, F.~Michler, J.~Fuchs, R.~Weigel, C.~Ostgathe, and A.~Koelpin, ``Radar-based heart sound detection,'' \emph{Scientific Reports}, vol.~8, no.~1, pp. 1--14, Jul. 2018.

\bibitem{nosrati2017high}
M.~Nosrati and N.~Tavassolian, ``{High-accuracy heart rate variability monitoring using Doppler radar based on Gaussian pulse train modeling and FTPR algorithm},'' \emph{IEEE Transactions on Microwave Theory and Techniques}, vol.~66, no.~1, pp. 556--567, Jan. 2017.

\bibitem{will2016instantaneous}
C.~Will, K.~Shi, F.~Lurz, R.~Weigel, and A.~Koelpin, ``Instantaneous heartbeat detection using a cross-correlation based template matching for continuous wave radar systems,'' in \emph{2016 IEEE Topical Conference on Wireless Sensors and Sensor Networks (WiSNet)}, pp. 31--34, Jan. 2016.

\bibitem{will2017advanced}
C.~Will, K.~Shi, R.~Weigel, and A.~Koelpin, ``Advanced template matching algorithm for instantaneous heartbeat detection using continuous wave radar systems,'' in \emph{2017 First IEEE MTT-S International Microwave Bio Conference (IMBIOC)}, pp. 1--4, May 2017.

\bibitem{sakamoto2015feature}
T.~Sakamoto, R.~Imasaka, H.~Taki, T.~Sato, M.~Yoshioka, K.~Inoue, T.~Fukuda, and H.~Sakai, ``Feature-based correlation and topological similarity for interbeat interval estimation using ultrawideband radar,'' \emph{IEEE Transactions on Biomedical Engineering}, vol.~63, no.~4, pp. 747--757, Apr. 2015.

\bibitem{mei2020fast}
Z.~Mei, Q.~Wu, Z.~Hu, and J.~Tao, ``{A fast non-contact vital signs detection method based on regional hidden Markov model in a 77GHz LFMCW radar system},'' in \emph{IEEE International Conference on Acoustics, Speech and Signal Processing (ICASSP)}, pp. 1145--1149, May 2020.

\bibitem{shi2021contactless}
K.~Shi, T.~Steigleder, S.~Schellenberger, F.~Michler, A.~Malessa, F.~Lurz, N.~Rohleder, C.~Ostgathe, R.~Weigel, and A.~Koelpin, ``{Contactless analysis of heart rate variability during cold pressor test using radar interferometry and bidirectional LSTM networks},'' \emph{Scientific Reports}, vol.~11, no.~1, pp. 1--13, Feb. 2021.

\bibitem{bechet2013non}
P.~Bechet, R.~Mitran, and M.~Munteanu, ``A non-contact method based on multiple signal classification algorithm to reduce the measurement time for accurately heart rate detection,'' \emph{Review of Scientific Instruments}, vol.~84, no.~8, p. 084707, Aug. 2013.

\bibitem{yamamoto2018non}
K.~Yamamoto, K.~Toyoda, and T.~Ohtsuki, ``{Non-contact heartbeat detection by MUSIC with discrete cosine transform-based parameter adjustment},'' in \emph{2018 IEEE Global Communications Conference (GLOBECOM)}, pp. 1--6, Dec. 2018.

\bibitem{yamamoto2019music}
K.~Yamamoto, K.~Toyoda, and T.~Ohtsuki, ``{MUSIC-based non-contact heart rate estimation with adaptive window size setting},'' in \emph{2019 41st Annual International Conference of the IEEE Engineering in Medicine and Biology Society (EMBC)}, pp. 6073--6076, Jul. 2019.

\bibitem{mercuri2018direct}
M.~Mercuri, Y.-H. Liu, I.~Lorato, T.~Torfs, F.~Wieringa, A.~Bourdoux, and C.~Van~Hoof, ``{A direct phase-tracking Doppler radar using wavelet independent component analysis for non-contact respiratory and heart rate monitoring},'' \emph{IEEE Transactions on Biomedical Circuits and Systems}, vol.~12, no.~3, pp. 632--643, Jun. 2018.

\bibitem{lv2021non}
W.~Lv, W.~He, X.~Lin, and J.~Miao, ``{Non-Contact Monitoring of Human Vital Signs Using FMCW Millimeter Wave Radar in the 120 GHz Band},'' \emph{Sensors}, vol.~21, no.~8, p. 2732, Oct. 2021.

\bibitem{liu2020vital}
Z.~Liu, Y.~Kong, X.~Zhang, J.~Wu, and W.~Lu, ``Vital sign extraction in the presence of radar mutual interference,'' \emph{IEEE Signal Processing Letters}, vol.~27, pp. 1745--1749, Sep. 2020.

\bibitem{sun2020remote}
L.~Sun, S.~Huang, Y.~Li, C.~Gu, H.~Pan, H.~Hong, and X.~Zhu, ``{Remote measurement of human vital signs based on joint-range adaptive EEMD},'' \emph{IEEE Access}, vol.~8, pp. 68\,514--68\,524, Apr. 2020.

\bibitem{duan2018non}
Z.~Duan and J.~Liang, ``{Non-contact detection of vital signs using a UWB radar sensor},'' \emph{IEEE Access}, vol.~7, pp. 36\,888--36\,895, Dec. 2018.

\bibitem{wang2021mmhrv}
F.~Wang, X.~Zeng, C.~Wu, B.~Wang, and K.~R. Liu, ``{mmHRV: Contactless heart rate variability monitoring using millimeter-wave radio},'' \emph{IEEE Internet of Things Journal}, vol.~8, no.~22, pp. 16\,623--16\,636, Nov. 2021.

\bibitem{wang2019noncontact}
P.~Wang, F.~Qi, M.~Liu, F.~Liang, H.~Xue, Y.~Zhang, H.~Lv, and J.~Wang, ``{Noncontact heart rate measurement based on an improved convolutional sparse coding method using IR-UWB radar},'' \emph{IEEE Access}, vol.~7, pp. 158\,492--158\,502, Oct. 2019.

\bibitem{wu2019person}
S.~Wu, T.~Sakamoto, K.~Oishi, T.~Sato, K.~Inoue, T.~Fukuda, K.~Mizutani, and H.~Sakai, ``Person-specific heart rate estimation with ultra-wideband radar using convolutional neural networks,'' \emph{IEEE Access}, vol.~7, pp. 168\,484--168\,494, Nov. 2019.

\bibitem{li2019standalone}
Y.~Li, Z.~Xia, and Y.~Zhang, ``{Standalone systolic profile detection of non-contact SCG signal with LSTM network},'' \emph{IEEE Sensors Journal}, vol.~20, no.~6, pp. 3123--3131, Mar. 2020.

\bibitem{zheng2020v2ifi}
T.~Zheng, Z.~Chen, C.~Cai, J.~Luo, and X.~Zhang, ``{V2ifi: In-vehicle vital sign monitoring via compact RF sensing},'' \emph{Proceedings of the ACM on Interactive, Mobile, Wearable and Ubiquitous Technologies (IMWUT)}, vol.~4, no.~2, pp. 1--27, Jun. 2020.

\bibitem{rabiner1969chirp}
L.~Rabiner, R.~W. Schafer, and C.~Rader, ``The chirp z-transform algorithm,'' \emph{IEEE Transactions on Audio and Electroacoustics}, vol.~17, no.~2, pp. 86--92, Jun. 1969.

\bibitem{widrow1975adaptive}
B.~Widrow, J.~R. Glover, J.~M. McCool, J.~Kaunitz, C.~S. Williams, R.~H. Hearn, J.~R. Zeidler, J.~E. Dong, and R.~C. Goodlin, ``Adaptive noise cancelling: Principles and applications,'' \emph{Proceedings of the IEEE}, vol.~63, no.~12, pp. 1692--1716, Dec. 1975.

\bibitem{ren2016phase}
L.~Ren, H.~Wang, K.~Naishadham, O.~Kilic, and A.~E. Fathy, ``{Phase-based methods for heart rate detection using UWB impulse Doppler radar},'' \emph{IEEE Transactions on Microwave Theory and Techniques}, vol.~64, no.~10, pp. 3319--3331, Oct. 2016.

\bibitem{mercuri2013analysis}
M.~Mercuri, P.~J. Soh, G.~Pandey, P.~Karsmakers, G.~A. Vandenbosch, P.~Leroux, and D.~Schreurs, ``Analysis of an indoor biomedical radar-based system for health monitoring,'' \emph{IEEE Transactions on Microwave Theory and Techniques}, vol.~61, no.~5, pp. 2061--2068, May 2013.

\bibitem{ahmed1974discrete}
N.~Ahmed, T.~Natarajan, and K.~R. Rao, ``Discrete cosine transform,'' \emph{IEEE Transactions on Computers}, vol. 100, no.~1, pp. 90--93, Jan. 1974.

\bibitem{bentley1994wavelet}
P.~M. Bentley and J.~McDonnell, ``Wavelet transforms: An introduction,'' \emph{Electronics \& Communication Engineering Journal}, vol.~6, no.~4, pp. 175--186, Feb. 1994.

\bibitem{tariq2011vital}
A.~Tariq and H.~Ghafouri-Shiraz, ``{Vital signs detection using Doppler radar and continuous wavelet transform},'' in \emph{Proceedings of the 5th European Conference on Antennas and Propagation (EUCAP)}, pp. 285--288, Apr. 2011.

\bibitem{zhao2017noncontact}
H.~Zhao, H.~Hong, L.~Sun, Y.~Li, C.~Li, and X.~Zhu, ``{Noncontact physiological dynamics detection using low-power digital-IF Doppler radar},'' \emph{IEEE Transactions on Instrumentation and Measurement}, vol.~66, no.~7, pp. 1780--1788, Jul. 2017.

\bibitem{schmidt2010segmentation}
S.~E. Schmidt, C.~Holst-Hansen, C.~Graff, E.~Toft, and J.~J. Struijk, ``{Segmentation of heart sound recordings by a duration-dependent hidden Markov model},'' \emph{Physiological Measurement}, vol.~31, no.~4, p. 513, Apr. 2010.

\bibitem{cardoso1998blind}
J.-F. Cardoso, ``Blind signal separation: Statistical principles,'' \emph{Proceedings of the IEEE}, vol.~86, no.~10, pp. 2009--2025, Oct. 1998.

\bibitem{stone2004independent}
J.~V. Stone, \emph{{Independent Component Analysis: A Tutorial Introduction}}.\hskip 1em plus 0.5em minus 0.4em\relax Cambridge, MA, USA: MIT Press, 2004.

\bibitem{weishaupt2018vital}
F.~Weishaupt, I.~Walterscheid, O.~Biallawons, and J.~Klare, ``{Vital sign localization and measurement using an LFMCW MIMO radar},'' in \emph{2018 19th International Radar Symposium (IRS)}, pp. 1--8, Jun. 2018.

\bibitem{liang2018improved}
X.~Liang, H.~Zhang, S.~Ye, G.~Fang, and T.~A. Gulliver, ``{Improved denoising method for through-wall vital sign detection using UWB impulse radar},'' \emph{Digital Signal Processing}, vol.~74, pp. 72--93, Mar. 2018.

\bibitem{dragomiretskiy2013variational}
K.~Dragomiretskiy and D.~Zosso, ``Variational mode decomposition,'' \emph{IEEE Transactions on Signal Processing}, vol.~62, no.~3, pp. 531--544, Feb. 2013.

\bibitem{zhang2014troika}
Z.~Zhang, Z.~Pi, and B.~Liu, ``Troika: A general framework for heart rate monitoring using wrist-type photoplethysmographic signals during intensive physical exercise,'' \emph{IEEE Transactions on Biomedical Engineering}, vol.~62, no.~2, pp. 522--531, Feb. 2014.

\bibitem{candes2006stable}
E.~J. Candes, J.~K. Romberg, and T.~Tao, ``Stable signal recovery from incomplete and inaccurate measurements,'' \emph{Communications on Pure and Applied Mathematics: A Journal Issued by the Courant Institute of Mathematical Sciences}, vol.~59, no.~8, pp. 1207--1223, Mar. 2006.

\bibitem{luo2022pso} 
X.~Luo, Y.~Yuan, S.~Chen, N.~Zeng, and Z.~Wang, ``{Position-transitional particle swarm optimization-incorporated latent factor analysis},'' \emph{IEEE Transactions on Knowledge and Data Engineering}, vol.~34, no.~8, pp.~3958--3970, Aug. 2022.

\bibitem{zeng2022pso} 
N.~Zeng, Z.~Wang, W.~Liu, H.~Zhang, K.~Hone, and X.~Liu, ``{A dynamic neighborhood-based switching particle swarm optimization algorithm},'' \emph{IEEE Transactions on Cybernetics}, vol.~52, no.~9, pp.~9290--9301, Sep. 2022.

\bibitem{zhang2019end}
W.~Zhang, Q.~Zhang, J.~Cheng, C.~Bai, and P.~Hao, ``End-to-end panoptic segmentation with pixel-level non-overlapping embedding,'' in \emph{2019 IEEE International Conference on Multimedia and Expo (ICME)}, pp. 976--981, Jul. 2019.

\bibitem{hochreiter1997long}
S.~Hochreiter and J.~Schmidhuber, ``Long short-term memory,'' \emph{Neural Computation}, vol.~9, no.~8, pp. 1735--1780, Nov. 1997.

\bibitem{IJSS-2021-06}
J.~Hu, C.~Jia, H.~Liu, X.~Yi, and Y.~Liu, ``{A survey on state estimation of complex dynamical networks},'' \emph{International Journal of Systems Science}, vol.~52, no.~16, pp.~3351--3367, Dec. 2021.

\bibitem{cheng2022deep} 
H.~Cheng, Z.~Wang, Z.~Wei, L.~Ma, and X.~Liu, ``{On adaptive learning framework for deep weighted sparse autoencoder: A multiobjective evolutionary algorithm},'' \emph{IEEE Transactions on Cybernetics. IEEE Transactions on Cybernetics}, vol.~52, no.~5, pp.~3221--3231, May 2022.

\bibitem{chen2020simple}
T.~Chen, S.~Kornblith, M.~Norouzi, and G.~Hinton, ``A simple framework for contrastive learning of visual representations,'' in \emph{International Conference on Machine Learning}, pp. 1597--1607, Feb. 2020.

\bibitem{jaiswal2020survey}
A.~Jaiswal, A.~R. Babu, M.~Z. Zadeh, D.~Banerjee, and F.~Makedon, ``A survey on contrastive self-supervised learning,'' \emph{Technologies}, vol.~9, no.~1, p.~2, Dec. 2020.

\bibitem{shi2020dataset}
K.~Shi, S.~Schellenberger, C.~Will, T.~Steigleder, F.~Michler, J.~Fuchs, R.~Weigel, C.~Ostgathe, and A.~Koelpin, ``A dataset of radar-recorded heart sounds and vital signs including synchronised reference sensor signals,'' \emph{Scientific Data}, vol.~7, no.~1, pp. 1--12, Feb. 2020.

\bibitem{schellenberger2020dataset}
S.~Schellenberger, K.~Shi, T.~Steigleder, A.~Malessa, F.~Michler, L.~Hameyer, N.~Neumann, F.~Lurz, R.~Weigel, C.~Ostgathe, and A.~Koelpin, ``A dataset of clinically recorded radar vital signs with synchronised reference sensor signals,'' \emph{Scientific Data}, vol.~7, no.~1, pp. 1--11, Sep. 2020.

\bibitem{yoo2021radar}
S.~Yoo, S.~Ahmed, S.~Kang, D.~Hwang, J.~Lee, J.~Son, and S.~H. Cho, ``Radar recorded child vital sign public dataset and deep learning-based age group classification framework for vehicular application,'' \emph{Sensors}, vol.~21, no.~7, p. 2412, Mar. 2021.

\bibitem{edanami2022medical}
K.~Edanami and G.~Sun, ``Medical radar signal dataset for non-contact respiration and heart rate measurement,'' \emph{Data in Brief}, vol.~40, p. 107724, Feb. 2022.

\bibitem{zhai2022contactless}
Q.~Zhai, X.~Han, Y.~Han, J.~Yi, S.~Wang, and T.~Liu, ``A contactless on-bed radar system for human respiration monitoring,'' \emph{IEEE Transactions on Instrumentation and Measurement}, vol.~71, pp. 1--10, Apr. 2022.

\bibitem{rong2019remote}
Y.~Rong and D.~W. Bliss, ``Remote sensing for vital information based on spectral-domain harmonic signatures,'' \emph{IEEE Transactions on Aerospace and Electronic Systems}, vol.~55, no.~6, pp. 3454--3465, May 2019.

\bibitem{le2019multivariate}
M.~Le and B.~Van~Nguyen, ``{Multivariate correlation of higher harmonics for heart rate remote measurement using UWB impulse radar},'' \emph{IEEE Sensors Journal}, vol.~20, no.~4, pp. 1859--1866, Oct. 2019.

\bibitem{da2019theoretical}
S.~D. Da~Cruz, H.-P. Beise, U.~Schr{\"o}der, and U.~Karahasanovic, ``A theoretical investigation of the detection of vital signs in presence of car vibrations and radar-based passenger classification,'' \emph{IEEE Transactions on Vehicular Technology}, vol.~68, no.~4, pp. 3374--3385, Apr. 2019.

\bibitem{IJSS-2021-03} 
J.~Mao, Y.~Sun, X.~Yi, H.~Liu, and D.~Ding, ``{Recursive filtering of networked nonlinear systems: A survey},'' \emph{International Journal of Systems Science}, vol.~52, no.~6, pp.~1110--1128, Apr. 2021.

\bibitem{park2021preclinical}
J.-Y. Park, Y.~Lee, R.~Heo, H.-K. Park, S.-H. Cho, S.~H. Cho, and Y.-H. Lim, ``{Preclinical evaluation of noncontact vital signs monitoring using real-time IR-UWB radar and factors affecting its accuracy},'' \emph{Scientific Reports}, vol.~11, no.~1, p. 23602, Dec. 2021.

\bibitem{lee2018novel}
Y.~Lee, J.-Y. Park, Y.-W. Choi, H.-K. Park, S.-H. Cho, S.~H. Cho, and Y.-H. Lim, ``{A novel non-contact heart rate monitor using impulse-radio ultra-wideband (IR-UWB) radar technology},'' \emph{Scientific Reports}, vol.~8, no.~1, p. 13053, Aug. 2018.

\bibitem{lee2020feasibility}
W.~H. Lee, Y.~Lee, J.~Y. Na, S.~H. Kim, H.~J. Lee, Y.-H. Lim, S.~H. Cho, S.~H. Cho, and H.-K. Park, ``Feasibility of non-contact cardiorespiratory monitoring using impulse-radio ultra-wideband radar in the neonatal intensive care unit,'' \emph{PLoS One}, vol.~15, no.~12, p. e0243939, 2020.

\bibitem{cardillo2021vital}
E.~Cardillo, C.~Li, and A.~Caddemi, ``Vital sign detection and radar self-motion cancellation through clutter identification,'' \emph{IEEE Transactions on Microwave Theory and Techniques}, vol.~69, no.~3, pp. 1932--1942, Sep. 2021.

\bibitem{lin2022broadband}
Y.-H. Lin, J.-H. Cheng, L.-C. Chang, W.-J. Lin, J.-H. Tsai, and T.-W. Huang, ``{A broadband MFCW agile radar concept for vital-sign detection under various thoracic movements},'' \emph{IEEE Transactions on Microwave Theory and Techniques}, vol.~70, no.~8, pp. 4056--4070, Jul. 2022.

\bibitem{IJSS-2021-05} 
L.~Zou, Z.~Wang, J.~Hu, Y.~Liu, and X.~Liu, ``{Communication-protocol-based analysis and synthesis of networked systems: Progress, prospects and challenges},'' \emph{International Journal of Systems Science}, vol.~52, no.~14, pp.~3013--3034, Oct. 2021.

\bibitem{IJNDI-2022-09} Q.~Zhang and Y.~Zhou, ``{Recent advances in non-Gaussian stochastic systems control theory and its applications},'' \emph{International Journal of Network Dynamics and Intelligence}, vol.~1, no.~1, pp.~111-119, Dec.~2022. 

\bibitem{IJSS-2021-08} 
H.~Geng, H.~Liu, L.~Ma, and X.~Yi, ``{Multi-sensor filtering fusion meets censored measurements under a constrained network environment: Advances, challenges and prospects},'' \emph{International Journal of Systems Science}, vol.~52, no.~16, pp.~3410--3436, Dec. 2021.

\bibitem{gouveia2022bio}
C.~Gouveia, D.~F. Albuquerque, P.~Pinho, and J.~Vieira, ``{Bio-radar cardiac signal model used for HRV assessment and evaluation using adaptive filtering},'' \emph{IEEE Transactions on Instrumentation and Measurement}, vol.~71, pp. 1--10, Jul. 2022.

\bibitem{sai2022cognitive}
Y.~P. Sai and L.~R. Kumari, ``{Cognitive assistant DeepNet model for detection of cardiac arrhythmia},'' \emph{Biomedical Signal Processing and Control}, vol.~71, p. 103221, Jan. 2022.

\bibitem{guan2022man}
R.~Guan, K.~L. Man, H.~Zhao, R.~Zhang, S.~Yao, J.~Smith, E.~G. Lim, and Y.~Yue, ``{MAN and CAT: Mix attention to nn and concatenate attention to YOLO},'' \emph{The Journal of Supercomputing}, pp. 1--29, Aug. 2022.

\bibitem{SSCE-2022-02} M.~C.~Liang, W.~Y.~Liu, Y.~H.~Wen and H.~Yang, ``{Segmentation and weight prediction of grape ear based on SFNet-ResNet18},'' \emph{Systems Science \& Control Engineering}, vol.~10, no.~1, pp.~722-732, 2022.

\bibitem{SSCE-2021-01} P.~Lu, B.~Song and L.~Xu, ``{Human face recognition based on convolutional neural network and augmented dataset},'' \emph{Systems Science \& Control Engineering}, vol.~9, no.~s2, pp.~29-37, 2021.

\bibitem{chen2022msdan} 
Y.~Chen, R.~Yang, M.~Huang, Z.~Wang, and X.~Liu, ``{Single-source to single-target cross-subject motor imagery classification based on multisubdomain adaptation network},'' \emph{IEEE Transactions on Neural Systems and Rehabilitation Engineering}, vol.~30, pp.~1992--2002, Jul. 2022.

\bibitem{wan2021review}
Z.~Wan, R.~Yang, M.~Huang, N.~Zeng, and X.~Liu, ``{A review on transfer learning in EEG signal analysis},'' \emph{Neurocomputing}, vol. 421, pp. 1--14, Jan. 2021.

\bibitem{wan2022eeg}
Z.~Wan, R.~Yang, M.~Huang, W.~Liu, and N.~Zeng, ``{EEG fading data classification based on improved manifold learning with adaptive neighborhood selection},'' \emph{Neurocomputing}, vol. 482, pp. 186-196, Apr. 2022.

\bibitem{yin2019self}
W.~Yin, X.~Yang, L.~Li, L.~Zhang, N.~Kitsuwan, R.~Shinkuma, and E.~Oki, ``{Self-adjustable domain adaptation in personalized ECG monitoring integrated with IR-UWB radar},'' \emph{Biomedical Signal Processing and Control}, vol.~47, pp. 75--87, Aug. 2019.

\bibitem{IJNDI-2022-04} M.~Szankin and A.~Kwasniewska, ``{Can AI see bias in X-ray images?},'' \emph{International Journal of Network Dynamics and Intelligence}, vol.~1, no.~1, pp.~48-64, Dec.~2022. 

\bibitem{xu2021cardiacwave}
C.~Xu, H.~Li, Z.~Li, H.~Zhang, A.~S. Rathore, X.~Chen, K.~Wang, M.-c. Huang, and W.~Xu, ``{CardiacWave: A mmWave-based scheme of non-contact and high-definition heart activity computing},'' \emph{Proceedings of the ACM on Interactive, Mobile, Wearable and Ubiquitous Technologies (IMWUT)}, vol.~5, no.~3, pp. 1--26, Sep. 2021.
  
\bibitem{li2019intelligent}
L.~Li, Y.~Shuang, Q.~Ma, H.~Li, H.~Zhao, M.~Wei, C.~Liu, C.~Hao, C.-W. Qiu, and T.~J. Cui, ``Intelligent metasurface imager and recognizer,'' \emph{Light: Science \& Applications}, vol.~8, no.~1, pp. 1--9, Oct. 2019.

\end{thebibliography}
\end{document}